\documentclass[12pt]{iopart}

\usepackage{graphicx}
\begin{document}

\title[Hidden Markov Models for the detection of hidden orders]{Statistical identification with hidden Markov models of large order splitting strategies in an equity market}

\author{Gabriella Vaglica$^1$, Fabrizio Lillo$^{1,2}$, Rosario N. Mantegna$^1$}

\address{$^1$ Dipartimento di Fisica e Tecnologie Relative, Universit\`a di Palermo,
Viale delle Scienze, I-90128, Palermo, Italy}
\address{$^2$ Santa Fe Institute, 1399 Hyde Park
Road, Santa Fe, NM 87501, USA}

\ead{lillo@unipa.it}

\begin{abstract}
Large trades in a financial market are usually split into smaller parts and traded incrementally over extended periods of time. We address these large trades as hidden orders.
In order to identify and characterize hidden orders we fit hidden Markov models to the time series of the sign of the tick by tick inventory variation of market members of the Spanish Stock Exchange. 
Our methodology probabilistically detects trading sequences, which are characterized by a net majority of buy or sell transactions. We interpret these patches of sequential buying or selling transactions as proxies of the traded hidden orders. We find that the time, volume and number of transactions size distributions of these patches are fat tailed. Long patches are characterized by a high fraction of market orders and a low participation rate, while short patches  have a large fraction of limit orders and a high participation rate. We observe the existence of a buy-sell asymmetry in the number, average length, average fraction of market orders and average participation rate of the detected patches. The detected asymmetry is clearly depending on the local market trend. We also compare the hidden Markov models patches with those obtained with the segmentation method used in Vaglica {\it et al.} (2008) and we conclude that the former ones can be interpreted as a partition of the latter ones.  
\end{abstract}

\maketitle

\newpage

\section{Introduction}

Financial markets are complex systems where many heterogeneous agents interact. Most financial markets are continuous double auction markets where market members are allowed to trade by submitting limit and market orders. When a limit order is submitted, an order indicating the willingness to buy (or sell) a given volume of a financial asset at a given price is placed inside the order book. The order book is essentially a list of unexecuted orders waiting for a matching order. Market members can also send market orders which result in immediate transactions at the best available price. Market members channel all orders to be processed in the market. They may act on behalf of a third party or on their own interest. Market members are heterogeneous with respect to many characteristics. Examples are their size (in volume transacted), their trading time horizons (ranging from less than a second to months) and their portfolio of clients (institutional, individuals, companies, foreign investors, etc).

A large body of financial literature has investigated the statistical regularities observed around trades of different classes of investors such as institutional investors \cite{Chan1995,Grinblatt1995,Nofsinger1999,Choe1999,Grinblatt2000}, individual investors   \cite{Nofsinger1999,Grinblatt2000,Gallagher2006,Barber2008,Grinblatt2009} or foreign investors \cite{Choe1999,Grinblatt2000}. 
In the investigation of the trading profile of institutional investors it has been recognized the necessity institutional investors have to split large orders into many smaller transactions
\cite{Chan1995,barra,almgrenrisk,Lillo2004,lillo05,Gallagher2006,Vaglica2008,Bouchaud2009,Moro2009}. In this paper, as already done in previous studies \cite{Lillo2004,lillo05,Vaglica2008,Bouchaud2009,Moro2009}, we address these large orders as {\it hidden orders}\footnote{Other names used in the literature are large trades, packages, or meta-orders.} to emphasize that their true size is typically not public information.  The strategic reasons for incremental execution of hidden orders were originally analyzed by Kyle \cite{kyle}, who developed a model for an insider trader who has information about future prices. Typically traders of large orders attempt to keep the true size of their orders secret in order to minimize the execution cost of the orders.  Consider for example a trader who wishes to buy a large number of shares of a company because she is expecting future price increase.  She of course wishes to buy shares at the lowest price possible.  On the other hand, as she demands a large amount of shares she will push the price up.  By executing her order incrementally she should be able to buy at least part of the order at a price not significantly affected by her action, therefore minimizing her impact on the price dynamics of the stock of interest and lowering her overall execution cost \cite{Bouchaud2009,Moro2009}. 
Some earlier works have provided empirical evidence of statistical regularities of hidden orders \cite{Chan1995,Gallagher2006}. The empirical part of these works usually rely on specialized databases providing information about the coded identity of a set of investors. Information about the class of investors to whom each investor belong to is typically also provided to the researchers. These specialized databases are often confidential and proprietary and therefore quite difficult to access for research purposes.  
 
More recently another approach to this kind of investigations has been proposed in the econophysics literature. In this approach the trading dynamics of the most active market members of a large financial market has been investigated to statistically detect the strategic behavior in the execution of large orders \cite{Vaglica2008},  specialization in the trading strategies \cite{Lillo2008f,Zovko2007}, herding profile  \cite{Lillo2008f}, and price impact of large orders broken up into many small trades \cite{Moro2009}. These studies have successfully detected heterogeneity of market members with respect to their size, trade horizon and trading profile specialization both at the Spanish stock market and at the London Stock Exchange. In Ref. \cite{Vaglica2008} large trades broken up into many smaller orders have been statistically reconstructed and their statistical properties characterized  whereas in Ref. \cite{Moro2009} the price impact of large orders characterized by a large fraction of market orders was found concave and approximately increasing as the square root  of order size. 

The statistical reconstruction of hidden orders is not unique. References \cite{Vaglica2008,Moro2009} adapted an algorithm \cite{Bernaola2001} originally designed to investigate biomedical time series to the segmentation of the inventory profile of market members. This algorithm  is quite robust against small local fluctuations which are distinct from the overall global trend. This choice was motivated by the assumption that, at a given time, a market member could not be considered acting only for a specific client and therefore small deviations from the major trend should be expected at any time and taken into account by the algorithm in the process of segmentation. A corollary of this assumption is that the detected hidden order refers to the hidden order of the most important client of the investigated market member.

In the present work we take a different perspective. Here we propose a statistical method for the detection of hidden orders which is more sensitive to the local persistence of the trading profile of market members than the one used in refs  \cite{Vaglica2008,Moro2009}. We achieve this goal by implementing a segmentation procedure based on hidden Markov models (HMM). Hidden Markov models \cite{rabiner} have been introduced in the late 1960s and studied since then. They are very rich and flexible in their mathematical structure and are used in several fields ranging from speech and image recognition to the detection of biologically relevant nucleotide subsequences. Hidden Markov models \cite{hamilton} have also been used in econometrics and finance. Some examples are in refs \cite{Goldfeld1973,Tyssedal1988,Hamilton1989,Ryden1998,Diebold2001}.  
To the best of our knowledge this article presents the first application of HMM to the inventory time series of market members in an equity market.

In this paper we use HMM\footnote{In \ref{appendix} we also consider one of the most used extension of HMMs called hidden semi Markov models  \cite{Ferguson1980}.} to detect large trading orders that are split into pieces and executed incrementally. This method is used to identify periods of time when the tick by tick inventory variation sign is in the same underlying states according to the fitted model. 
We call the detected periods {\it patches}. Since market members act  simultaneously as brokers for many clients, it may be frequent that in a patch not all the transactions have the same sign. In the present study, we are mainly interested in directional patches, i.e. patches where the trader consistently buys or sells a large amount of shares. Our working hypothesis is that each of these patches contains at least one hidden order. The patches detected are analyzed with respect to their characteristics (time duration, number of transactions and volume exchanged), their liquidity profile (relative composition of limit and market orders) and their size compared to the total volume exchanged (participation rate of the patch transactions with respect to all transactions occurring in the market during the same time interval). 
The analysis is done separately for buying and selling directional patches. By keeping this two kind of orders separate we are able to detect that the statistical properties of patches detected by the HMM are showing a buy-sell asymmetry which depends on the state of the market.
We also compare the results obtained with HMM with previous results obtained with the more global segmentation algorithm of Ref.s \cite{Vaglica2008,Moro2009}.

The paper is organized as follows. In section 2 we recall the main properties of HMMs. Section 3 discusses the data investigated and the kind of methodology used to detect patches. In section 4 we present the results obtained when extracting the most likely HMM whereas in section 5 we investigate the statistical properties of detected patches. Section 6 discusses the asymmetry observed in buy and sell patches detected in time periods characterized by a different market state and section 7 discusses the comparison of the HMM methodology with the methodology based on a different segmentation algorithm recently used  \cite{Vaglica2008}. Section 8 concludes. In \ref{appendix} we discuss how patches characterized by a length with a power-law distribution are reconstructed by an HMM and we present a comparison of the results obtained by HMM and hidden semi Markov models.

\section{Hidden Markov models}

The identification of hidden orders is a typical problem of regime switching with hidden states. In regime switching models, the parameters of the model change abruptly from time to time and the time series is organized in consecutive patches (regimes) each one characterized by a set of model parameters. There is a vast amount of literature of regime switching models \cite{hamilton} and HMMs are among the most used ones. Even if one considers the time series of the trading activity of a single investor, it is not immediate to identify the set of consecutive transactions that originated from a unique trading decision and the boundary between different regimes. This problem is even more serious if, as in the analysis presented in this paper, we do not have the inventory variation of an individual investor but rather we have the one of a market member or a broker whose trading activity is generated by many investors. In this last case the identification of different regimes must be performed with statistical methods, such as the HMM.

An HMM  is a generalization of a Markov chain in which the states of the chain are not observable.
More specifically, consider a system described by a discrete time series and where the system can be in one of $N$ states. When the system is in a given state, it emits a signal with a given probability distribution of possible observations. The transition of the system from a state to another is described by a Markov chain. It is often the case that only the observations of the signal are observable while the underlying states are hidden. Such a system is described by an HMM.

More formally, the elements that completely characterize an HMM are the following \cite{rabiner}:
\begin{enumerate}
    \item $N$, the number of states in the model, $S = \{S_1,S_2,...,S_N\}$.
    \item $M$, the number of distinct observation symbols in each state. The observations correspond to the output of the
    signal. We denote the individual symbols as $V =\{v_1,v_2,...,v_M\}$. 
    \item The state transition probability matrix $ {\mathbf A}=\{a_{i,j}\}$ where
    \begin{equation}\label{markchain}
      a_{ij} = P[q_{t+1}= S_j|q_t = S_i], \hskip2cm 1\leq i,j \leq N
    \end{equation}
    where $q_{t+1}$ and $q_t$ are the state of the system at time $t+1$ and $t$, respectively.
    \item The observation symbol probability distribution in state $j$, ${\mathbf B}= \{b_j(k)\}$, called emission probabilities, where

     \begin{eqnarray}
      &&  b_j(k)= P[v_k \hskip0.3cm \textrm{at} \hskip0.3cm t| q_t =S_j], \hskip1.3cm 1\leq j \leq N \hskip1.2cm    1\leq k \leq M \label{Bpar}
     \end{eqnarray}
    \item The initial state distribution $\pi = \{\pi_i\}$ where
     \begin{eqnarray}
     && \pi_i = P[q_1 = S_i], \hskip3.2cm 1\leq i \leq N
    \end{eqnarray}
\end{enumerate}

The specification of the parameters $N$ and $M,$ and the specification of the three probability measures ${\mathbf A}$, ${\mathbf B}$ and $\pi$ completely define a given HMM (that we address simply as $\lambda(A, B,\pi)$ as in \cite{rabiner}).

Given an empirical time series of observed symbols $O=\{v_{t_1},v_{t_2},... \}$ one can ask for the best HMM able to explain it. To this end one has to compute $P(O|\lambda)$, the probability of observations' time series $O$ given a specific HMM $\lambda$. Then one finds the optimal HMM parameters by maximizing this probability.  This maximum likelihood estimation can be done either through an iterative procedure (Baum-Welch reestimation procedure) or by using gradient techniques \cite{rabiner}. 

After having found the best model $\lambda$ able to fit the data, one can ask for the ``optimal" sequence of hidden states $Q=\{S_{t_1},S_{t_2},...\}$ which best explains the observation sequence $O$ given the model $\lambda$. While the solution of the first problem (i.e. to find the best model $\lambda$) is unique,
for the second problem there are several optimality criteria. For example, one possibility is to ask for the sequence of states $Q$ which are individually most likely (this optimality criterion maximizes the expected number of correct individual states) or one can ask for the state sequence that maximizes the expected number of correct pairs of states or triplets and so on. One can also ask for the single best state sequence. An algorithm exists to solve this last problem and it is called the Viterbi algorithm \cite{Viterbi1967,Forney1973}. In the following we will search for the best sequence of states which are individually most likely.

HMMs are parametric models and therefore assume a given structure of the data. One of the most important structure is the Markovian assumption of the transition between states. As shown in \ref{appendix},  in an HMM the distribution of patch length is exponential. Therefore the HMM poses a serious constraint to the shape of the distribution of patch length. The distribution of real hidden order size is not known but there are indications that it might be characterized by long tails. Recently, by using a non parametric segmentation algorithm, it has been suggested that the tail of the distribution of hidden order size is described by a power law \cite{Vaglica2008}. Thus it seems that fitting the inventory variation with an HMM might pose an unrealistic structure to the patch size distribution. To verify the nature and degree of the constraints imposed by our methodology to the detection of patches characterized by a long tailed distribution we have performed extensive numerical simulations which are discussed in the \ref{appendix}.
As detailed in \ref{appendix} the patches detected with the HMM are quite close to the ones obtained with the more computationally intensive method called hidden semi Markov model (HSMM) which in numerical simulations and in real data is detecting patches distributions having approximately the same distribution as the one detected with HMM. Therefore it seems that the use of the more sophisticated (and probably more appropriate) HSMM does not change significantly the conclusions we draw by using the HMM. Given the intense computational work needed to apply the HSMM to the whole dataset and the requirement of long time series for the input sequence of this method, in the present study we have chosen to use HMM.

\section{Data and methods}\label{methods}

Our database of the electronic market SIBE (Sistema de Interconexi\'on Burs\'atil Electr\'onico) allows us to follow each on-book transaction performed by all the market members registered at Spanish Stock Exchange (BME, Bolsas y Mercados Espa\~noles).  In 2004 the BME was the eighth in the world in market capitalization.
We consider market members transactions for four highly capitalized stocks, specifically  Banco Bilbao Vizcaya Argentaria (BBVA),  Repsol (REP), Banco Santander Central Hispano (SAN), and Telef\'onica (TEF). Our database covers the time period 2001-2006. However in most of the following analysis we will restrict our attention to the period 2003-2006. The reason of this choice is that in 2001-2002 we observe a behavior which is markedly different from the one observed in 2003-2006. Since we show results pooled across different years, we restrict our attention to the years 2003-2006 which are more homogeneous. However in Section \ref{20012002} we discuss a buy-sell asymmetry providing a justification of the observation that some results change when we investigate the years 2001-2002.   
 
At BME market members are local and foreign credit entities and investment firms which are members of the stock exchange and are the only firms entitled to trade.
In fact, orders to buy and sell are entered into the market only through members of the stock market. Approximately, 75\% of them are major financial institutions and 25\% are established security dealers. Both kind of members may trade on their own behalf and/or on behalf of other individuals and institutions which are not members of the market. It is important to stress that market members are not necessarily quoted companies (stocks) but rather are the only institutions entitled to trade stocks directly. In this paper we consider only the most active market members defined by the criterion that each market member makes at least $1,000$ transactions per year and is active at least $200$ days per year in each of the six investigated years. The number of investigated market members is $43$ for BBVA, $37$ for REP, $46$ for SAN, and $52$ for TEF. These market members perform approximately $80\%$ of the on-book transactions and are responsible of approximately $78\%$ of the on-book volume (see Table \ref{summary} for the precise values for each investigated stock).

We fit an HMM  to the time series of the sign of the transactions rather than to the signed volume as in ref.s \cite{Vaglica2008,Moro2009}. The reason of this choice is that in this way we can fit an HMM without making any parametric assumption on the transaction volume distribution. The sign of a transaction is $+1$ if the investigated member buys and $-1$ if she sells. 
In the fit, performed by using a R software package named {\it hmm.discnp}, we set the number of hidden states of HMM equal to $N=3$  having in mind that they could correspond to a buy state, a sell state, and a state without a well defined trading direction. As we will see below, in most cases the fitted emission probabilities for these states confirm this interpretation. Therefore we fit each time series of the sign of the transactions of a market member with an HMM with $N=3$ and $M=2$. Since the detected patches are rather short and in order to detect possible slow time dependencies of the model parameters, we decide to fit the HMM and to apply the state reconstruction algorithm to each year separately. Since we cannot present the result of all the $(43+37+46+52)\times 4 =712$ fitted models, in the following we will show the results obtained for the 4 stocks and 4 years lumped together. As an example of the single asset behavior we also show the result for one specific year and stock, which is TEF in 2004.

\section{Fitting of hidden Markov models}

We use the expected maximization algorithm to perform the maximum likelihood fit of an HMM to the time series of transaction sign for each market member. Specifically, for each market member trading a stock in a year, we extract the most likely HMM. Since the observed variable is binary ($\pm1$), the emission probabilities are characterized by assigning, for example, the buy probability (+1) for each HMM state. We label the three states as states 1, 2, and 3. In the three states the average (across market members) buy emission probability is $ (0.92 \pm 0.07,  0.50 \pm0.14,  0.075 \pm 0.05)$  for TEF in 2004, while for the pooled sample it is $(0.95\pm 0.05,  0.51 \pm 0.17,  0.06 \pm0.05)$, where  errors are standard deviations. 
These values show that state 1 is characterized by a buy probability  very close to one, while state 3 is characterized by a buy probability very close to zero.

Transaction in state 1 are therefore preferentially buy transactions whereas in state 3 transactions are preferentially sell transactions. State 2 is characterized by an approximately balanced number of buy and sell transactions.  Therefore we call state 1 {\it buy state} and state 3 {\it sell state}. These two states are collectively termed {\it directional states}. In contrast the buy probability in state 2 is centered at $1/2$ and we call state 2 {\it neutral state}, i.e. a state describing a prolonged activity of the market member without a specified direction to buy or sell. It is worth noting that the standard deviation of the buy probability in state 2 is quite large. This shows that for some market members the state 2 is associated with a buy probability significantly different from $1/2$. In other words these market members lack a proper neutral state. We do not have an interpretation of this fact. It is finally worth noting that this effect is more evident in 2001 and 2002.

We then consider the transition probability matrices  ${\mathbf A}$ of the fitted HMM for each market member trading a stock in one year.
The average (across market members) transition probability matrix for TEF in 2004 is 
\begin{eqnarray}
 {\mathbf A}_{TEF04}=
\left(\begin{array}{ccc}
 0.89 \pm 0.06 &      0.07 \pm 0.05  & 0.04 \pm 0.05 \\
 0.06 \pm 0.04 &      0.87 \pm  0.08  &  0.07 \pm 0.07 \\
 0.04 \pm 0.05 &      0.06 \pm 0.04   &  0.90 \pm  0.07\\
 \end{array}\right)
\end{eqnarray}
while for the whole sample of 4 stocks and 4 years is
\begin{eqnarray}
 {\mathbf A}_{pool}=
\left(\begin{array}{ccc}
0.89 \pm 0.07    &   0.06 \pm 0.05 &   0.04 \pm 0.06 \\
0.07 \pm 0.05 &    0.85 \pm 0.07 &    0.08 \pm 0.06 \\
0.04 \pm 0.06 &    0.07 \pm 0.05 &    0.89 \pm 0.08 \\
 \end{array}\right)
\end{eqnarray}
where again errors are standard deviations. The analysis of the transition probability matrix shows that the diagonal elements, i.e. the conditional probabilities that the market member remains in the same state are quite large, indicating the presence of long periods in which the market member persists in the same  state. Moreover the off diagonal elements are on average very similar indicating that when the system is in a given state, it has a high probability to remain in its state but, if it changes state, it has almost equal probability to jump to either one of the other two states. Finally the fitted transition probability matrices are quite consistent across the four stocks and the four years.

\section{Detecting patches and their statistical properties}

\begin{figure}[ptb]
\begin{center}
\includegraphics[scale=0.4]{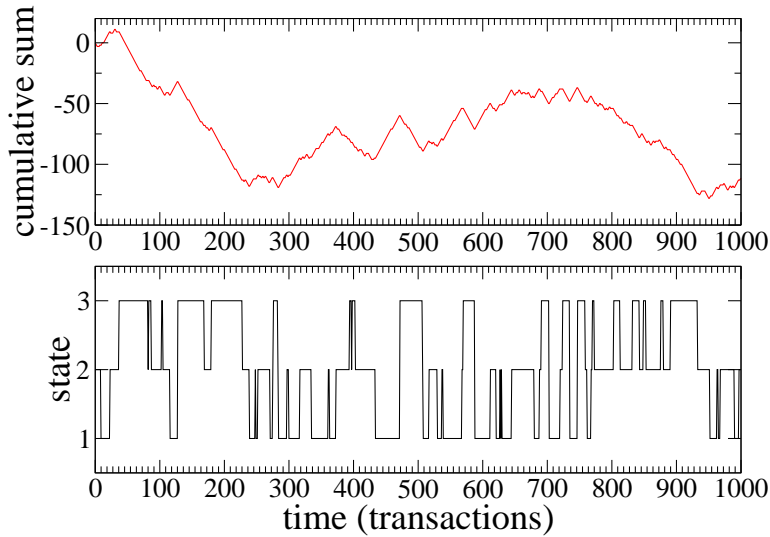}
\end{center}
\caption{Top panel. Cumulative sum of the transaction sign of a sequence of $1,000$ transactions of the market member Interdin Bolsa, Barcelona (code $9409$) when trading TEF. Clear trends or ramps are seen. Bottom panel. Corresponding most probable state according to the fitted HMM. State 1, 2, and 3 are the buy, neutral, and sell state, respectively.}
\label{9409}
\end{figure}

After the fitting of the HMM to the transaction sign time series, we calculate the most probable hidden state underlying each transaction for all the investigated market members.  We call these intervals of transactions patches of sequential buying or selling and interpret them as proxies of the traded hidden orders. To give an example of this reconstruction, in the top panel of Figure~\ref{9409} we show the cumulative sum of a sequence of $1,000$ transaction signs made by market member Interdin Bolsa, Barcelona (code $9409$) when trading TEF. The cumulative sum of the transaction sign shows the presence of ramps corresponding to time periods when the market member was preferentially buying or selling.
The identified HMM states in the bottom panel confirm our interpretation of the model states.

\subsection{Summary statistics of HMM patches}

We summarize some statistical properties of the detected HMM patches in Table \ref{summary}. We first observe that the number of patches is very large being more than $10^5$ per year and per stock.  The unconditional mean length of the directional patches is around $10$ transactions, but the standard deviation is significantly larger suggesting a non exponential length distribution. For neutral patches the mean length is a bit smaller but the standard deviation is again much larger than the mean. The distributional properties of the length of the patches will be investigated in  more detail in Section \ref{statprop}. Around $30\%$ of the patches are longer than $10$ transactions and this subsample will be investigated in depth in the following. This threshold was also used in Ref.  \cite{Vaglica2008,Moro2009}. The mean length of these long patches is between 20 and 30 transactions. The typical number of patches of a market member in a year is $1,000$.

\begin{center}
\begin{table*}[t!]
\caption{Summary statistics of  the patches detected in the time series of transaction signs of active market members. Data refer to the four investigated stocks for the period 2003-2006. Line 1 gives the total number of transactions for the considered period. Line 2 gives the number of active market members selected.
Line 3 and line 4 give the percentage of on-book transactions and on-book volume in which at least one of the investigated market members is involved,  respectively. Lines 5 and 6 give the total number of
patches in directional states  and in the neutral state, respectively. Lines 7 and 8 is the total number of
patches with at least 10 transactions in directional and neutral states, respectively. Lines 9-12 give the mean and standard deviation (sd) of patch length (in transactions) obtained by considering directional and neutral patches of all the selected market members. Lines 13-16 give the mean and standard deviation of patch length (in transactions) obtained by considering directional and neutral patches with at least 10 transactions of all the selected market members.}
\small{ 
\begin{tabular}{|l|lcccc|}\hline\hline
      &     &    BBVA       &   REP    & SAN   &     TEF  \\
        &   &  &    &   &                           \\
1&total number of transactions                          &4,413,563 & 3,289,239 & 4,914,738 & 5,940,651  \\
2&number of active market members                      & 43 & 37 & 46 & 52  \\
3&\% of transactions of active market members  & 77.1 & 75.1 & 82.0 & 83.9 \\
4&\% of volume (shares) of active market members  & 72.5 & 74.1 & 79.7 & 83.8 \\
5&total number of directional patches              & 319,057 & 228,647 &301,691 & 478,175  \\
6&total number of neutral patches                  & 192,308 & 137,252 & 192,236 & 267,160 \\
7&total number of directional patches ($N \ge 10$) & 87,758 & 66,737 & 107,073 & 133,208  \\
8&total number of neutral patches ($N \ge 10$)     & 42,792 & 30,656 & 56,807 & 72,005  \\
9&mean length of directional patches & 10.51 & 11.12 & 12.55 & 9.73  \\
10&sd of length of directional patches & 31.11 & 22.14 &30.78 & 19.70  \\
11&mean length of neutral patches & 7.41 & 7.09 & 9.78 & 8.62  \\
12&sd of length of neutral patches & 13.23 & 9.57 & 20.54 & 14.22  \\
13&mean length of directional patches  ($N \ge 10$) & 27.42 & 28.99 & 26.91 & 24.01  \\
14&sd of length of directional patches ($N \ge 10$) & 55.77 & 34.86 & 48.37 & 33.11  \\
15&mean length of neutral patches ($N \ge 10$) & 20.97 & 19.34 & 24.27 & 22.20  \\
16&sd of length of neutral patches ($N \ge 10$) & 23.02 & 14.00 & 33.37 & 21.95  \\
 \hline
\end{tabular}
}
\label{summary}
\end{table*}
\end{center}

Our results are based on the investigation of the time series of the signs of the transactions of each market member. In this way we discard any information concerning the volume (in shares or euros) of the transactions. Since volume is an highly fluctuating quantity \cite{gopi00,lillo05,eisler06}, it is important to verify that there is consistency between the interpretation of states 1 and 3 as directional states and the sign of the net volume exchanged in these states. Each patch is characterized by the total volume purchased $V_{buy}$ and the total volume sold $V_{sell}$. We also define $V_{tot}=V_{buy}+V_{sell}$ and we use Euro to measure volume. Analogously we also define $N_{tot}=N_{buy}+N_{sell}$ where  $N_{buy}$ is the number of buy transaction and $N_{sell}$ is the number of sell transactions present in the patch. 

\begin{figure}[ptb]
\begin{center}
\includegraphics[scale=0.27]{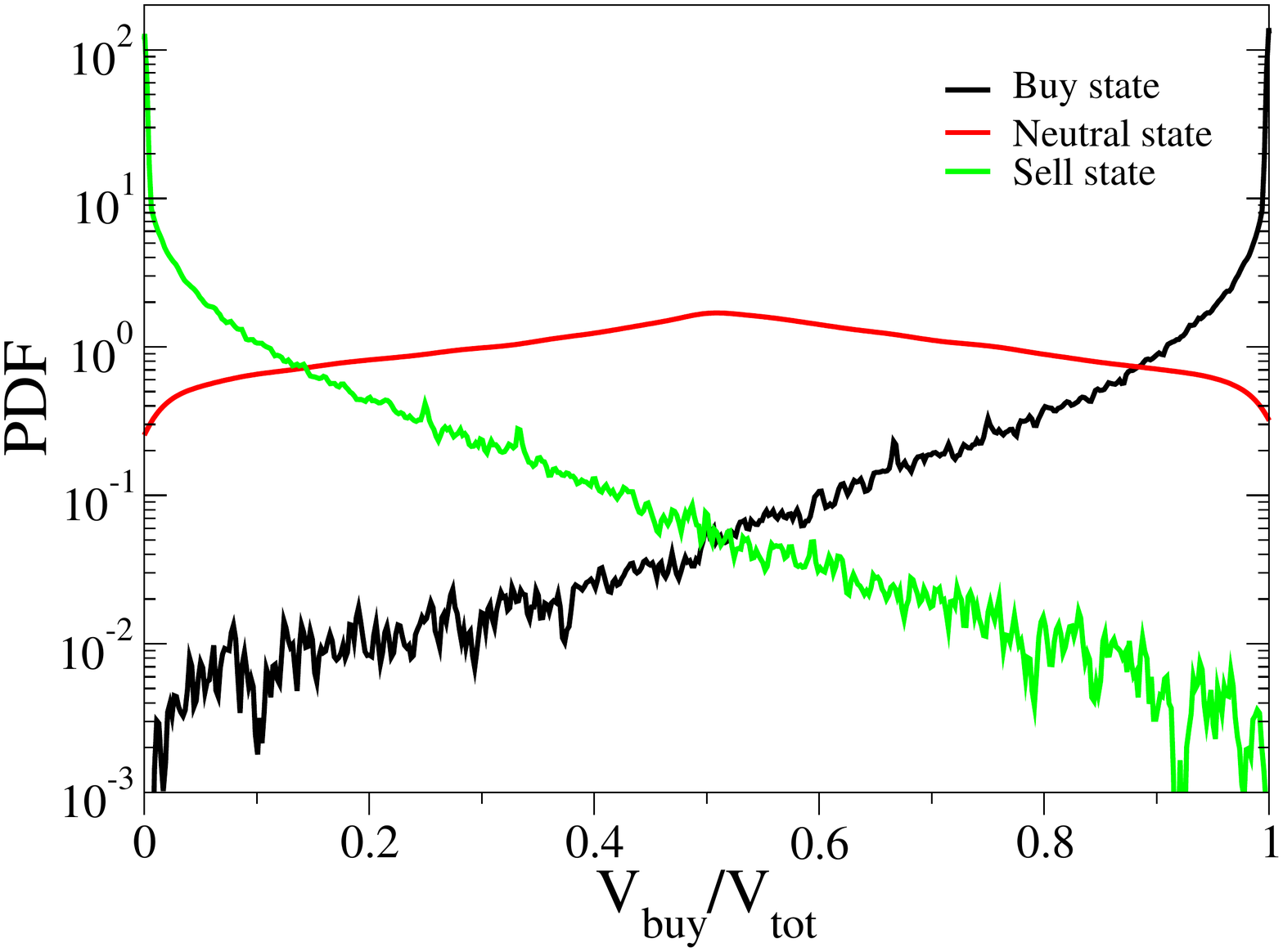}
\includegraphics[scale=0.27]{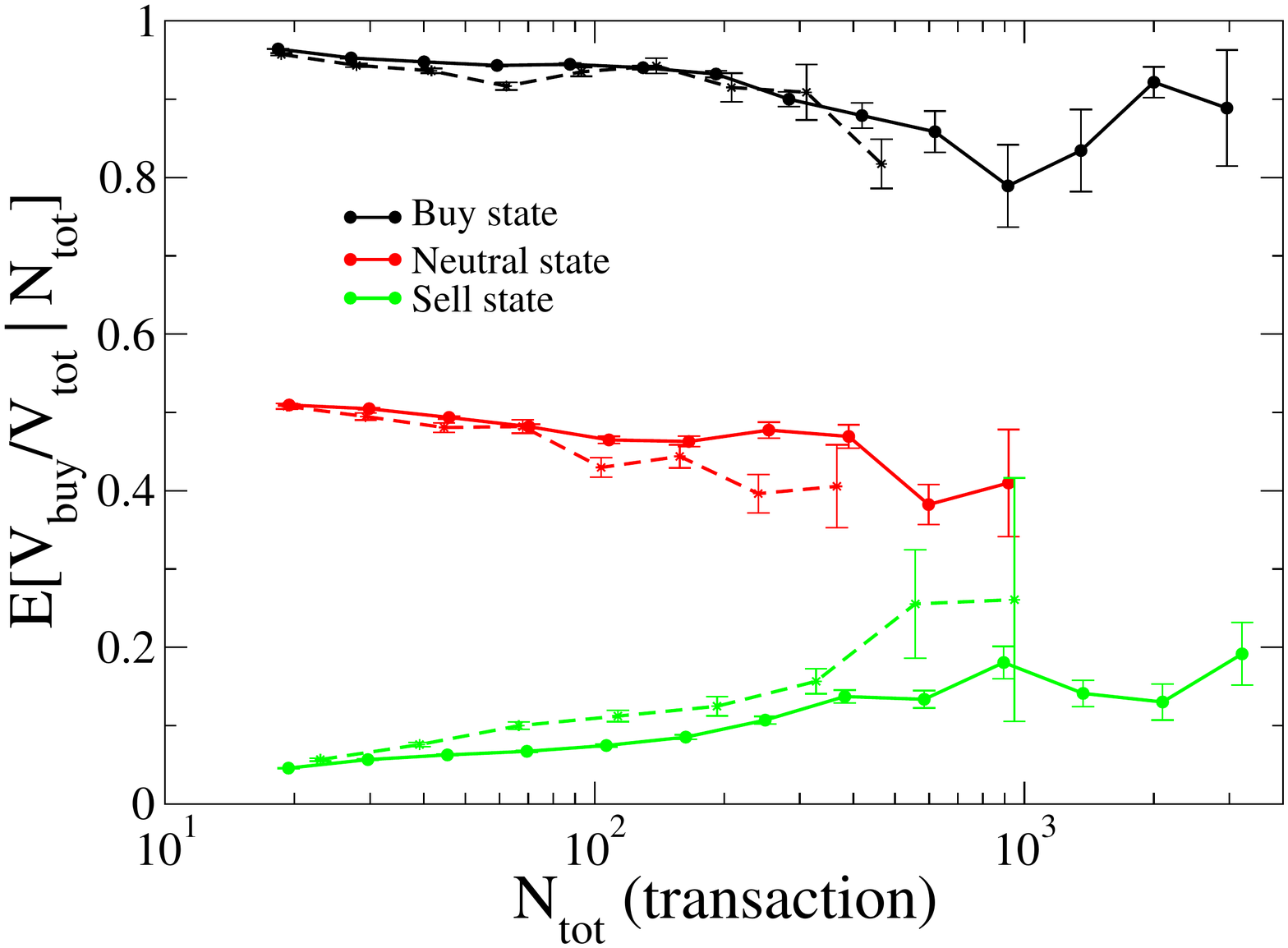}
\end{center}
\caption{Left panel. Probability density function of the ratio $V_{buy}/V_{tot}$ for the patches in the buy (black), neutral (red), and sell (green) state. We consider only patches with at least $10$ transactions for the pooled sample.  Right panel. Expected value of the ratio $V_{buy}/V_{tot}$ conditional to the patch length $N$ (in transactions) for the patches of TEF in 2004 (dashed lines) and for the pooled sample (solid lines).}
\label{vfrac}
\end{figure}

For each patch we compute the ratio $V_{buy}/V_{tot}$, and, in order to avoid discretization errors due to small number of transactions, we consider only patches with at least $10$ transactions. The average ratio for these patches is $0.97\pm0.09$, $0.5\pm 0.2$, and $0.04\pm0.09$ for patches in state buy, neutral, and sell, respectively. In the left panel of Figure \ref{vfrac} we show the PDF of the ratio for the patches in the three states.  The figure shows that the buy and sell state display a distribution of the ratio  $V_{buy}/V_{tot}$ which is very narrowly peaked close to 1 and zero, respectively. On the other hand this ratio is more broadly distributed for the neutral state. This result confirms the interpretation of state 1 and 3 as directional states, not only in number of transactions but also in volume. The left panel of Figure \ref{vfrac} considers all the patches with at least 10 transactions. The relation between our interpretation of the states and the ratio $V_{buy}/V_{tot}$ holds also if one measures the ratio conditional to the patch length. The right panel of Fig. \ref{vfrac} shows the mean ratio $V_{buy}/V_{tot}$ conditional to the patch length measured in total number of transactions. Long directional states tend to be slightly less pure, i.e. roughly $10\%$ of the volume in a long buy patch is associated to sell transactions. However this fluctuation is relatively small and we conclude that our interpretation of states 1 and 3 as buy and sell states respectively, remains valid even when one considers the amount of volume exchanged.

\subsection{Liquidity and participation rate of HMM patches}

\begin{figure}[ptb]
\begin{center}
\includegraphics[scale=0.27]{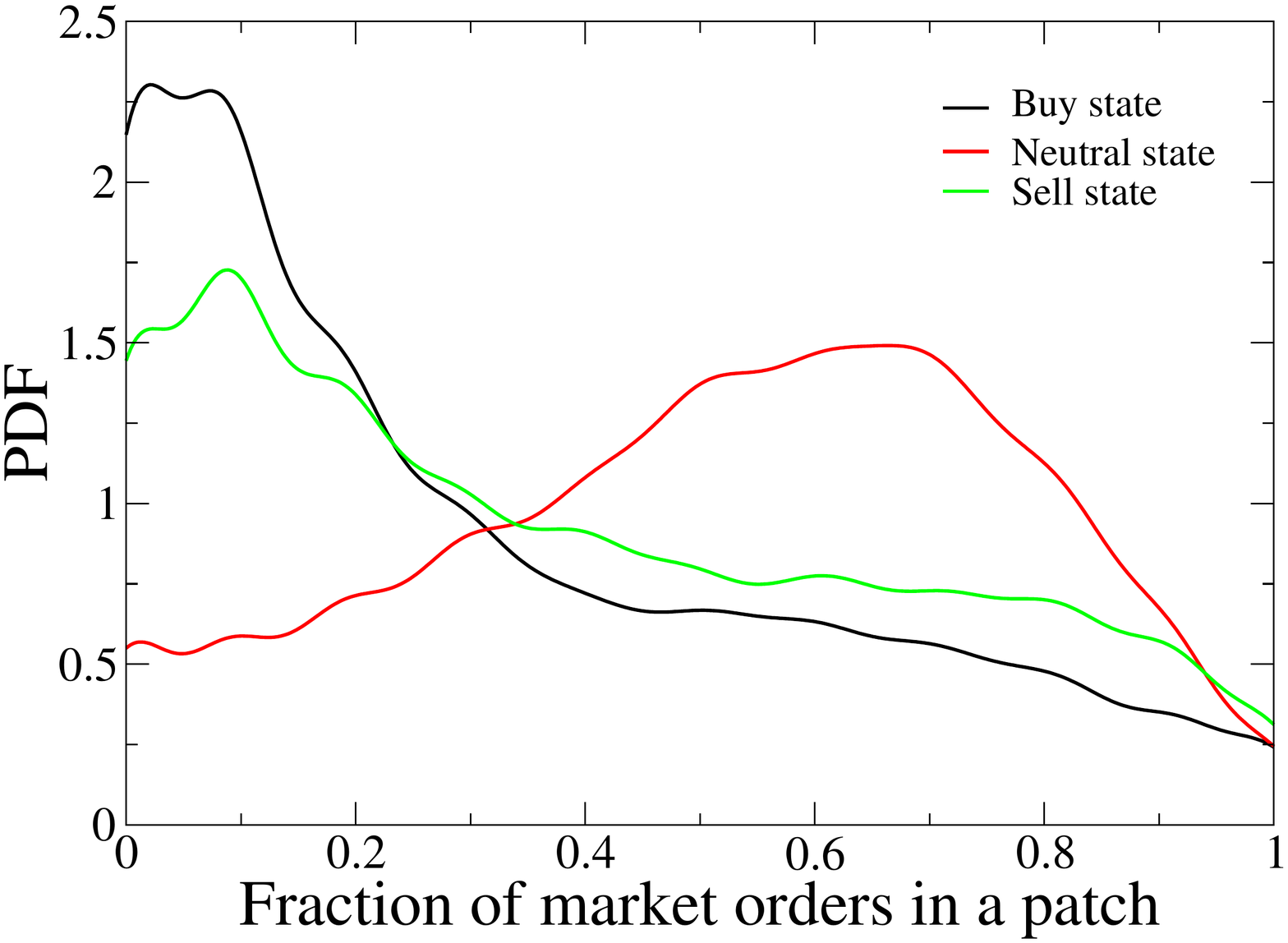}
\includegraphics[scale=0.27]{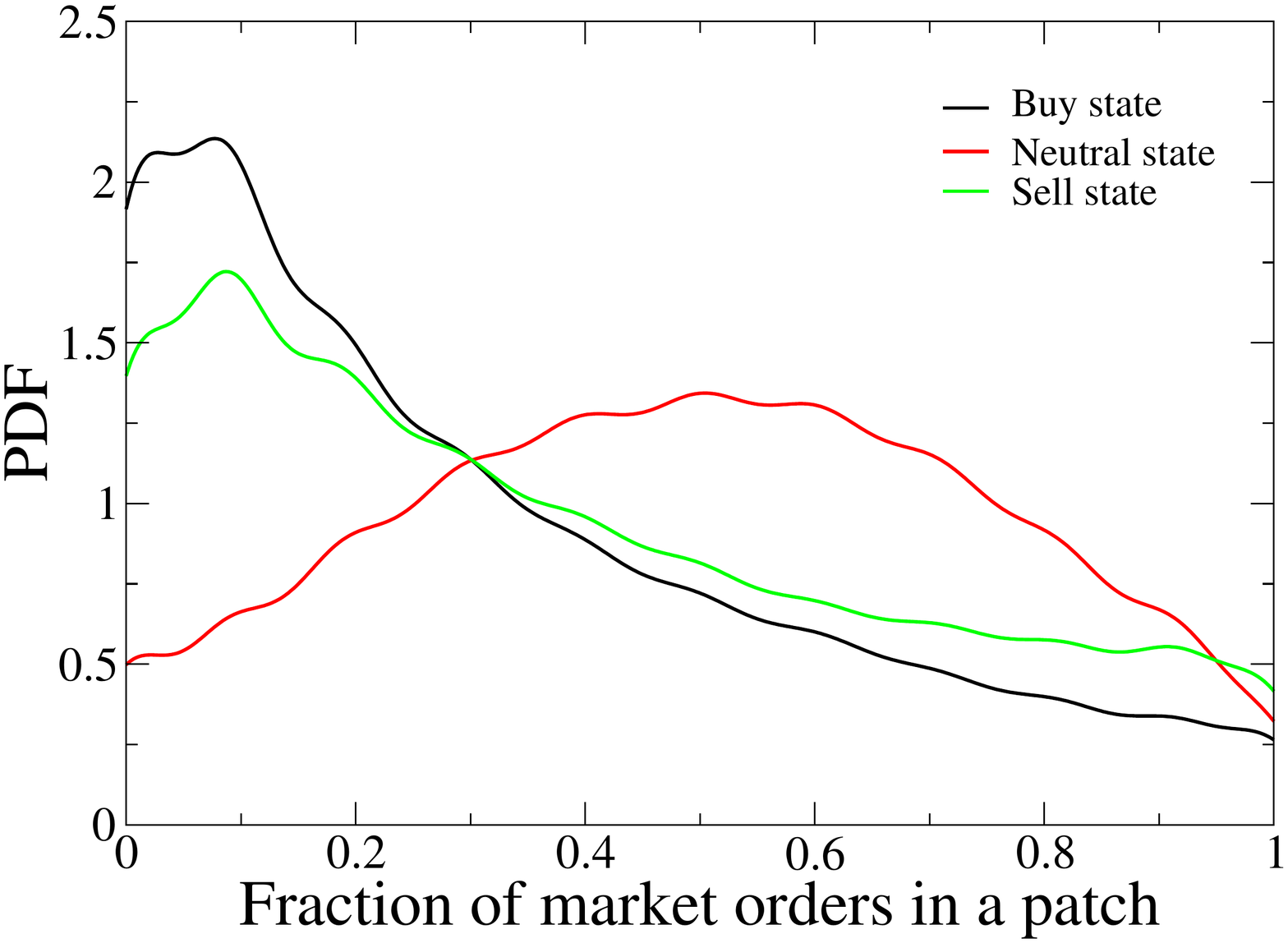}
\end{center}
\caption{Probability density function of the fraction of market orders of patches in the buy (black), neutral (red), and sell (sell) state for TEF in 2004 (left panel) and for the pooled sample (right panel). We consider only patches with at least $10$ transactions.}
\label{mopdf}
\end{figure}

Our working hypothesis is that HMM patches are related to the strategic behavior of traders, To provide empirical support to our hypothesis it is important to investigate the properties of the patches concerning the liquidity  properties and the participation rate of the patch. Each transaction in a electronic double auction market  happens between a limit order and an market order. The use of limit and market orders usually indicates different trading strategies. Market orders are placed by more impatient traders who want their orders to be executed as soon as possible, while traders using limit orders are more patient and are willing to wait for a transaction in order to get a better price. In our database we do not have direct information on who initiated the transaction, and therefore we use the Lee and Ready algorithm \cite{leeready} to infer if the initiator of the transaction, i.e. the agent who placed the market order, was the buyer or the seller. 

We investigate the fraction of market orders in the detected HMM patches. Figure \ref{mopdf} shows the probability density function of the fraction of market orders in a patch for the three states detected by the HMM. As done before, to avoid discretization problems we consider only patches made of at least $10$ transactions. The patches in state 2 show a fraction of market orders which is broadly distributed and the density function is roughly centered around $50\%$. On the contrary, directional patches show a clear preference for a low fraction of market orders. Moreover the figure shows that there is a difference between patches in buy and sell state. Buy patches tend to have a smaller proportion of market orders than sell patches. Patches in 2001-2002 show the opposite behavior.  

\begin{figure}[ptb]
\begin{center}
\includegraphics[scale=0.27]{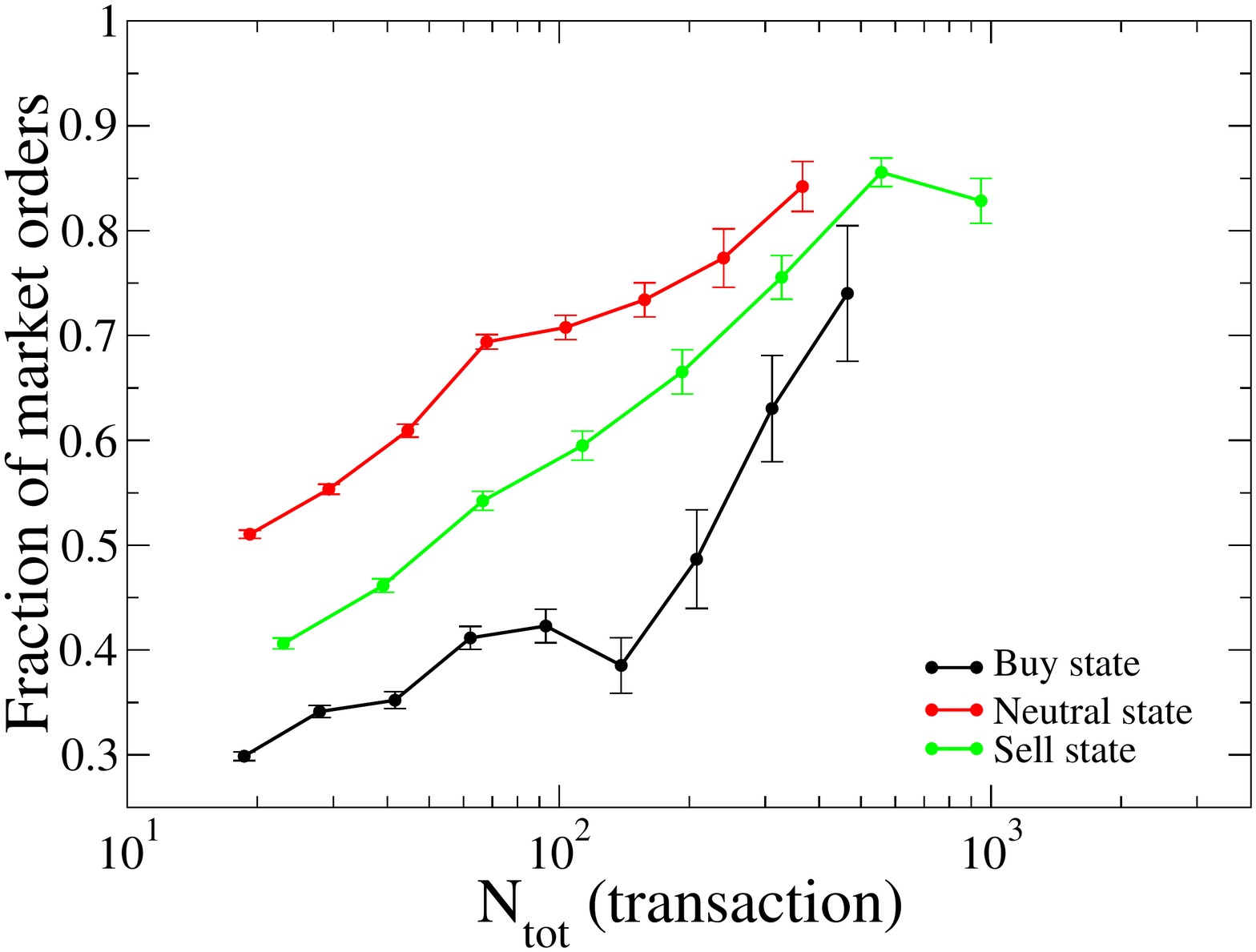}
\includegraphics[scale=0.27]{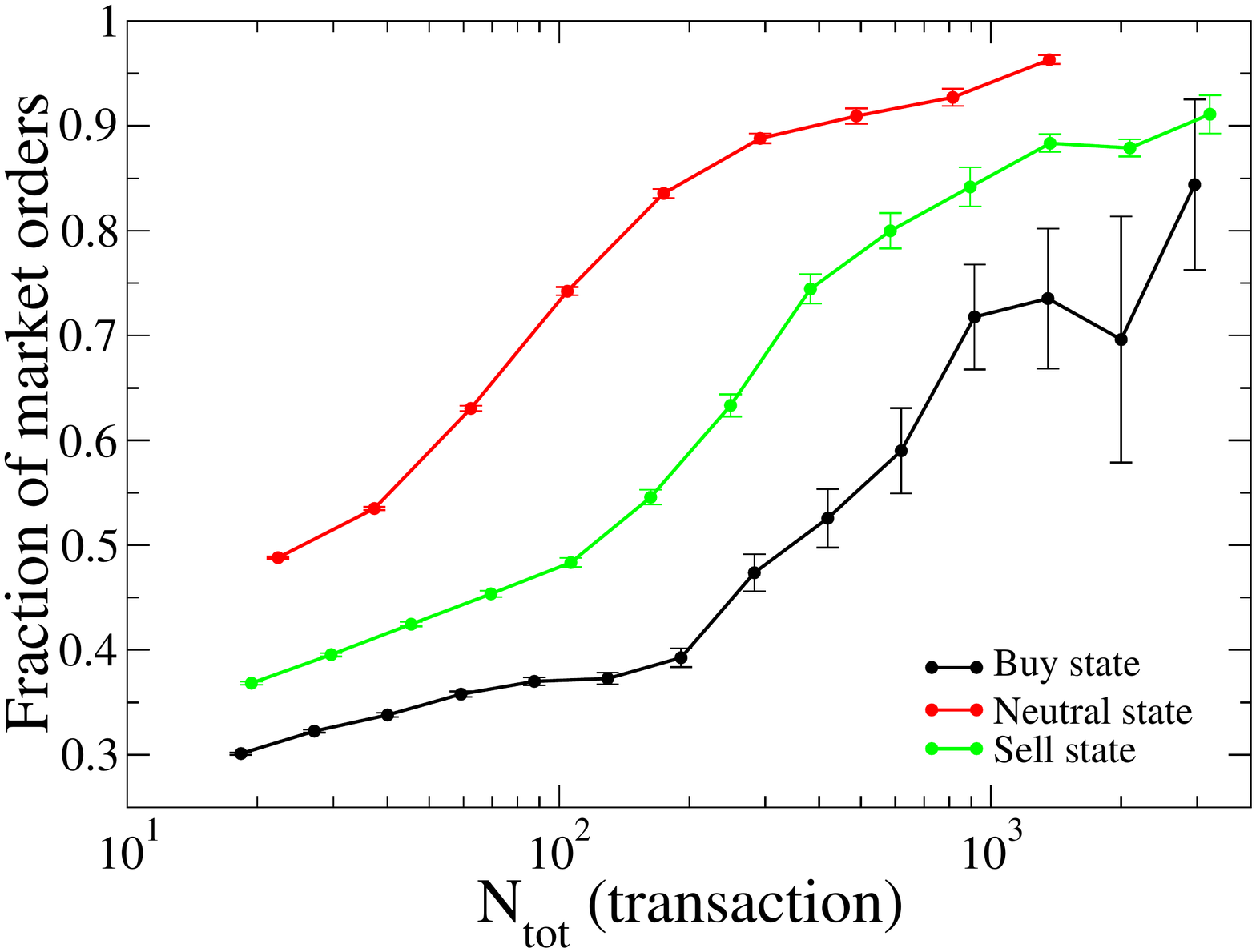}
\end{center}
\caption{Expected value of the fraction of market orders in a patch in the buy (black), neutral (red), and sell (green) state conditional to the patch length for TEF in 2004 (left panel) and for the pooled sample (right panel). Error bars are standard errors.}
\label{length-mo}
\end{figure}

The above result considers all the patches independently of their length. To investigate the dependency of the fraction of market orders from the patch length, we show the mean fraction of market orders conditional to the length of the patch in Figure \ref{length-mo}. The figure shows that for a given length, neutral patches are richer in market orders than patches in the sell state, which in turn are richer in market orders than patches in the buy state, confirming the unconditional result shown in Figure \ref{mopdf}. Again, the patches in 2001-2002 display the opposite buy-sell behavior. 
A possible explanation for the difference between the neutral and the directional states is market impact. It has been recently shown that an hidden order made of market orders has typically a positive impact, while an {\it executed} hidden order made of limit orders has typically a negative \footnote{See Ref.  \cite{Moro2009} for a discussion of this apparent paradox.} market impact \cite{Moro2009}. Thus if a trader wants to execute a buy or a sell hidden order, the use of limit orders will lower the impact when compared to the use of market order (of course, the use of limit orders does not give certainty of execution because the price could move in opposite direction). Therefore the low fraction of market orders in directional patches could be the consequence of a strategic choice of minimizing price impact. The asymmetry of behavior between buy and sell patches will be discussed in more detail in Section \ref{20012002}. 
Finally, figure \ref{length-mo} shows that the fraction of market orders in a patch increases with the patch length.

\begin{figure}[ptb]
\begin{center}
\includegraphics[scale=0.4]{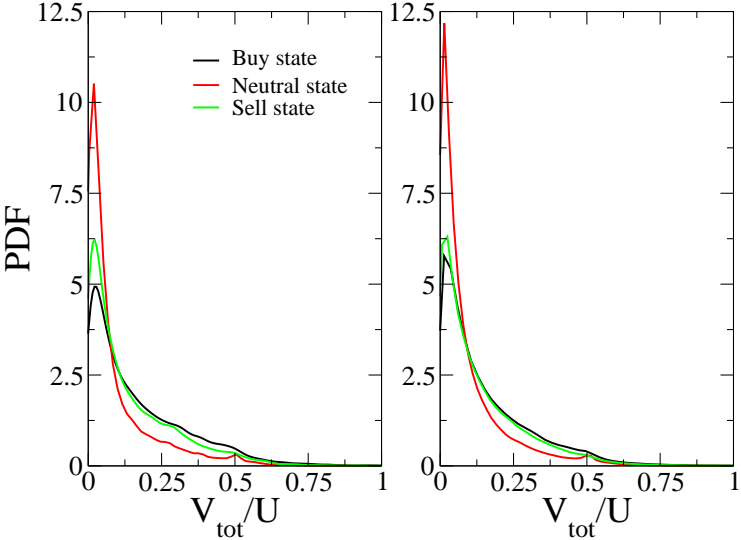}
\end{center}
\caption{Probability density function of the participation rate of patches in the buy (black), neutral (red), and sell (green) state  for TEF in 2004 (left panel) and the pooled sample (right panel).}
\label{aggressiveness-pdf}
\end{figure}

\begin{figure}[ptb]
\begin{center}
\includegraphics[scale=0.4]{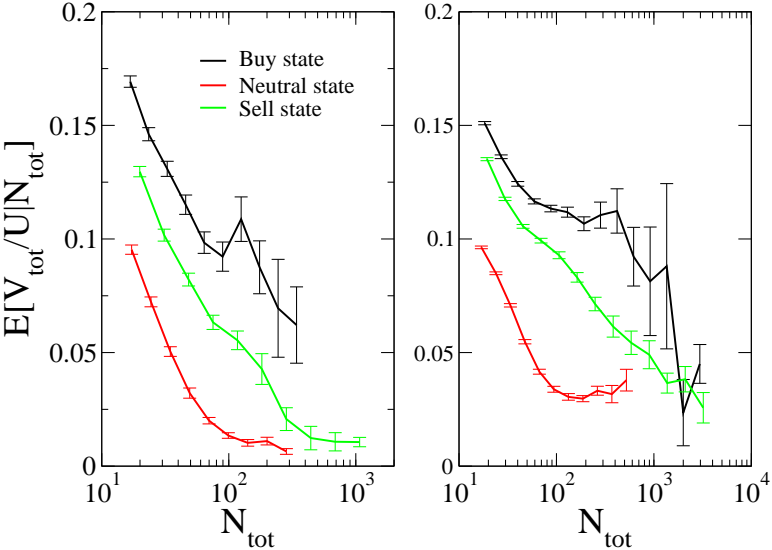}
\end{center}
\caption{Expected value of the participation rate conditional to the length of the patch in the buy (black), neutral (red), and sell (green) state for TEF in 2004 (left panel) and the pooled sample (right panel). Error bars are standard errors.}
\label{aggressiveness}
\end{figure}

\begin{figure}[ptb]
\begin{center}
\includegraphics[scale=0.27]{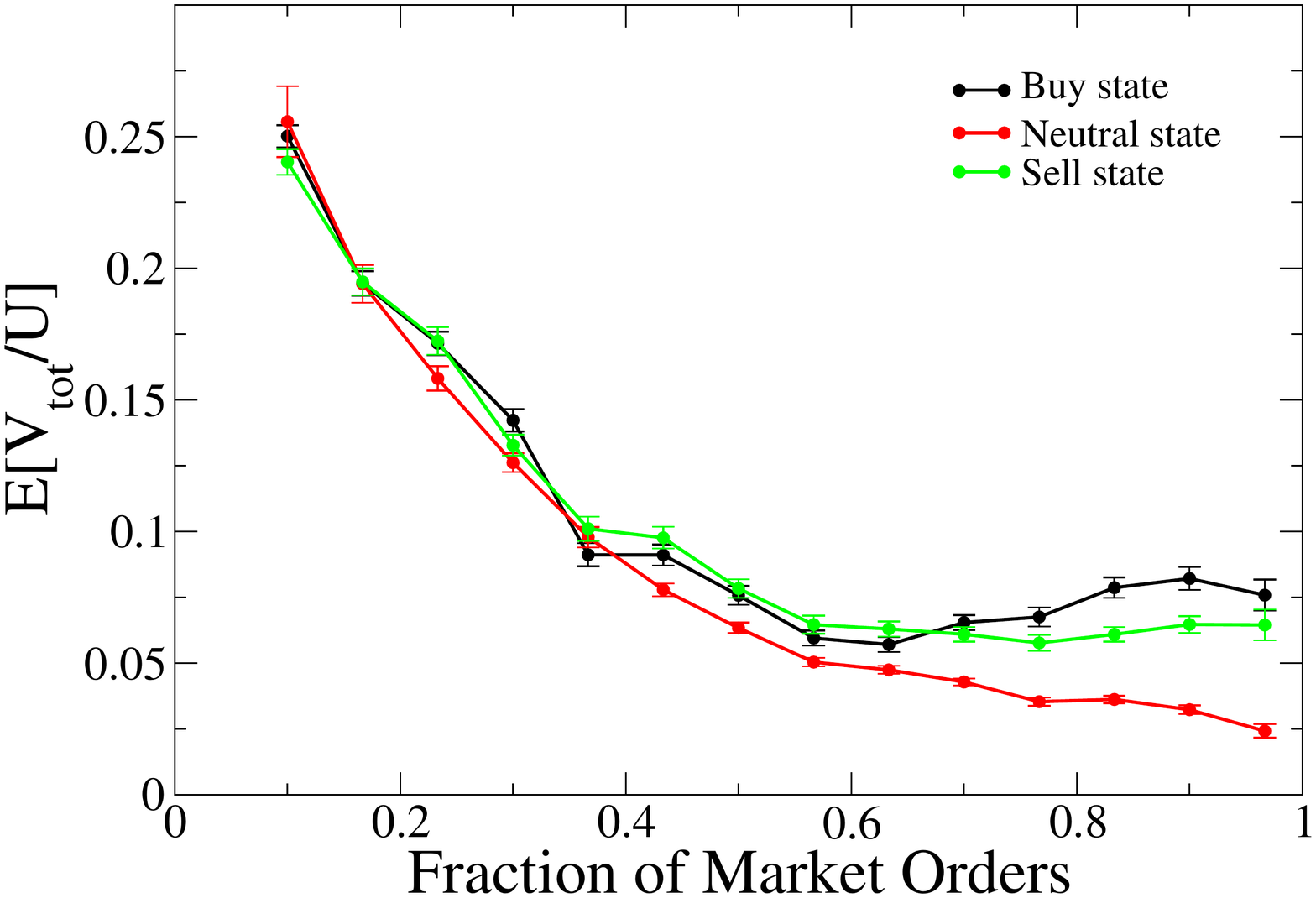}
\includegraphics[scale=0.27]{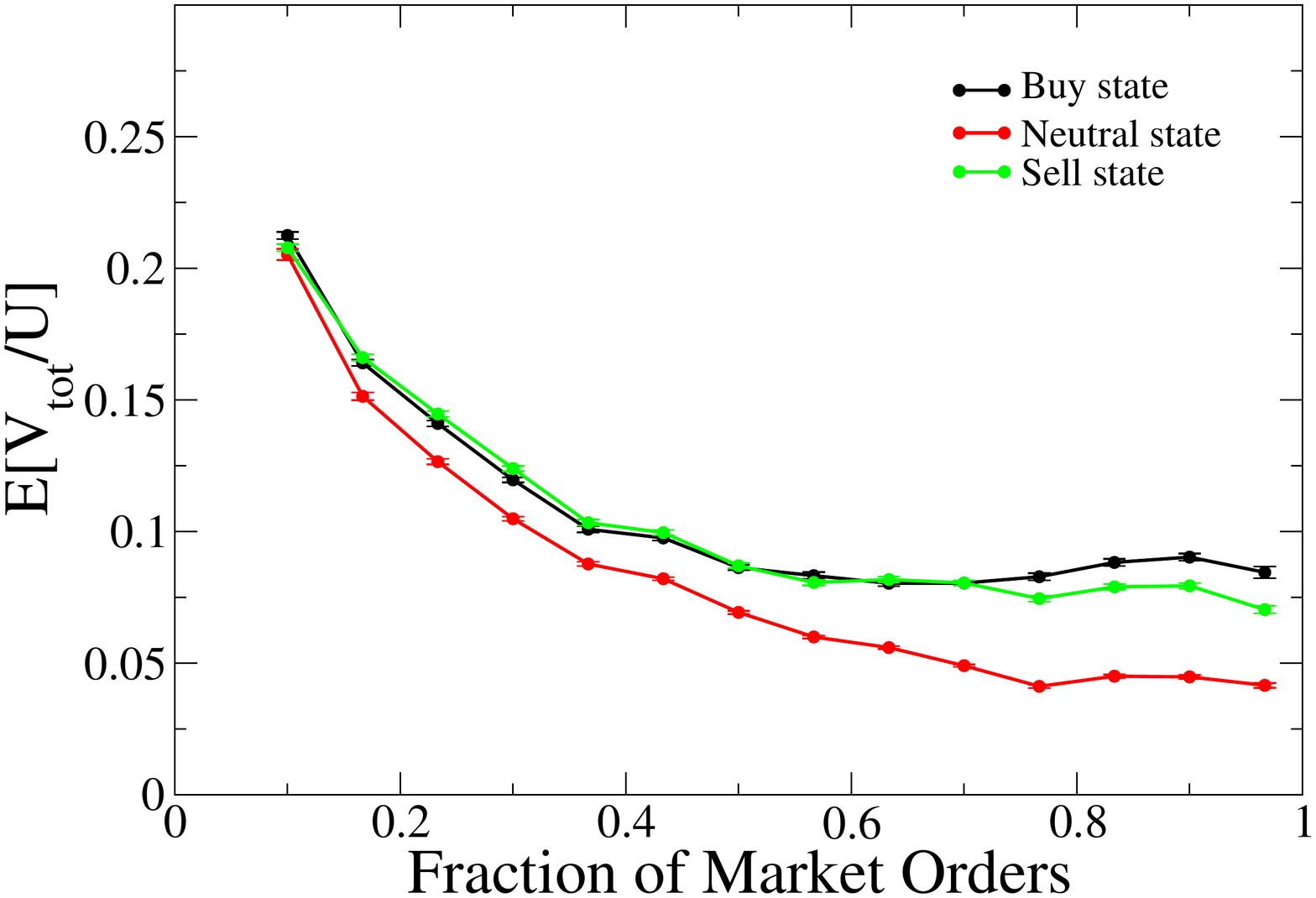}
\end{center}
\caption{Expected value of the participation rate conditional to the fraction of market orders of patches in the buy (black), neutral (red), and sell (sell) state for TEF 2004 (left panel) and the pooled sample (right panel). Error bars are standard errors.}
\label{aggressivenessmo}
\end{figure}

The other parameter characterizing the strategy behind hidden order placement is the participation rate. We define the participation rate of a patch as $V_{tot}/U$, where $V_{tot}$ is the total volume exchanged by the patch measured in Euros and $U$ is the total volume also measured in Euros exchanged in the traded stock during the execution of the patch. In Figure \ref{aggressiveness-pdf} we show the probability density function of the participation rate for the three states. Again we find a different behavior for the neutral and for the directional states. The neutral patches have a smaller participation rate than the directional patches. This result also holds when conditioning on the order length. Figure  \ref{aggressiveness} shows the mean value of the participation rate conditional to the patch length. We see that for a given patch length buy patches have on average a larger participation rate than sell patches, which in turn have a larger participation rate than neutral patches. As before in 2001-2002 the buy-sell asymmetry of behavior of directional patches is inverted. The participation rate decreases with the length of the patch and this is expected given that a large order is typically executed with a smaller participation rate in order to avoid detection by other traders and in order to decrease the impact cost. 
In Figure  \ref{aggressivenessmo} we consider the mean participation rate conditional to the fraction of market orders in a patch. There we find that this is a decreasing function, in contrast with what found in Ref. \cite{Moro2009} where patches detected with the more global segmentation algorithm were investigated. 

In summary, we conclude that large directional patches are made mainly of market orders and have a small participation rate, while small directional patches have a larger fraction of limit orders and a larger participation rate. The buy-sell asymmetry conforms to this pattern. In fact in the period 2003-2006  buy patches have a smaller fraction of market orders and a larger participation rate than sell patches.  In section \ref{20012002} we show that this asymmetry may be associated to the trend of the price of the traded stock.

\subsection{Statistical properties of HMM patch length size}\label{statprop}

\begin{figure}[thpb]
 \centering
 \vskip 1cm
 \includegraphics[width=8cm]{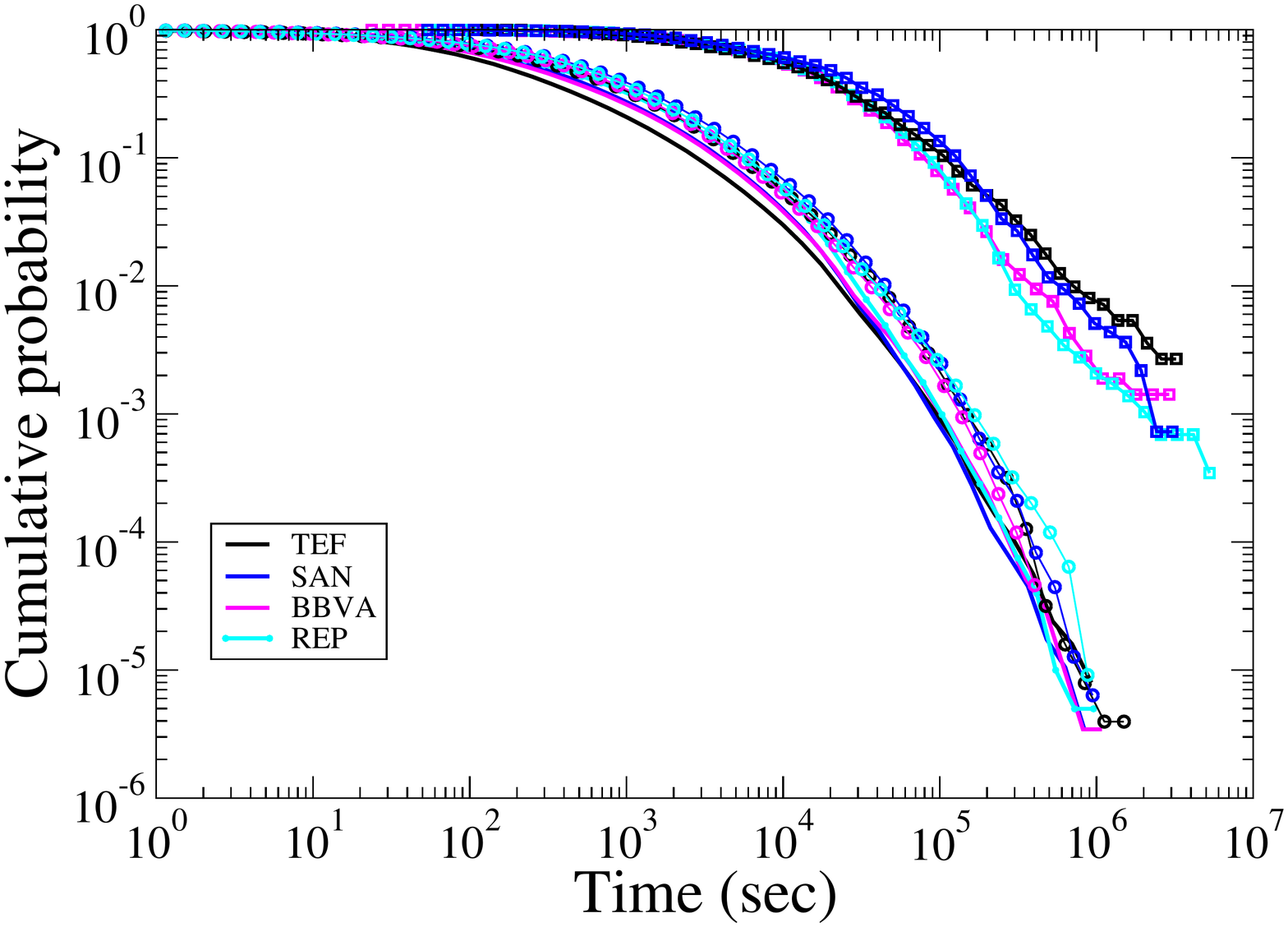}
 \includegraphics[width=8cm]{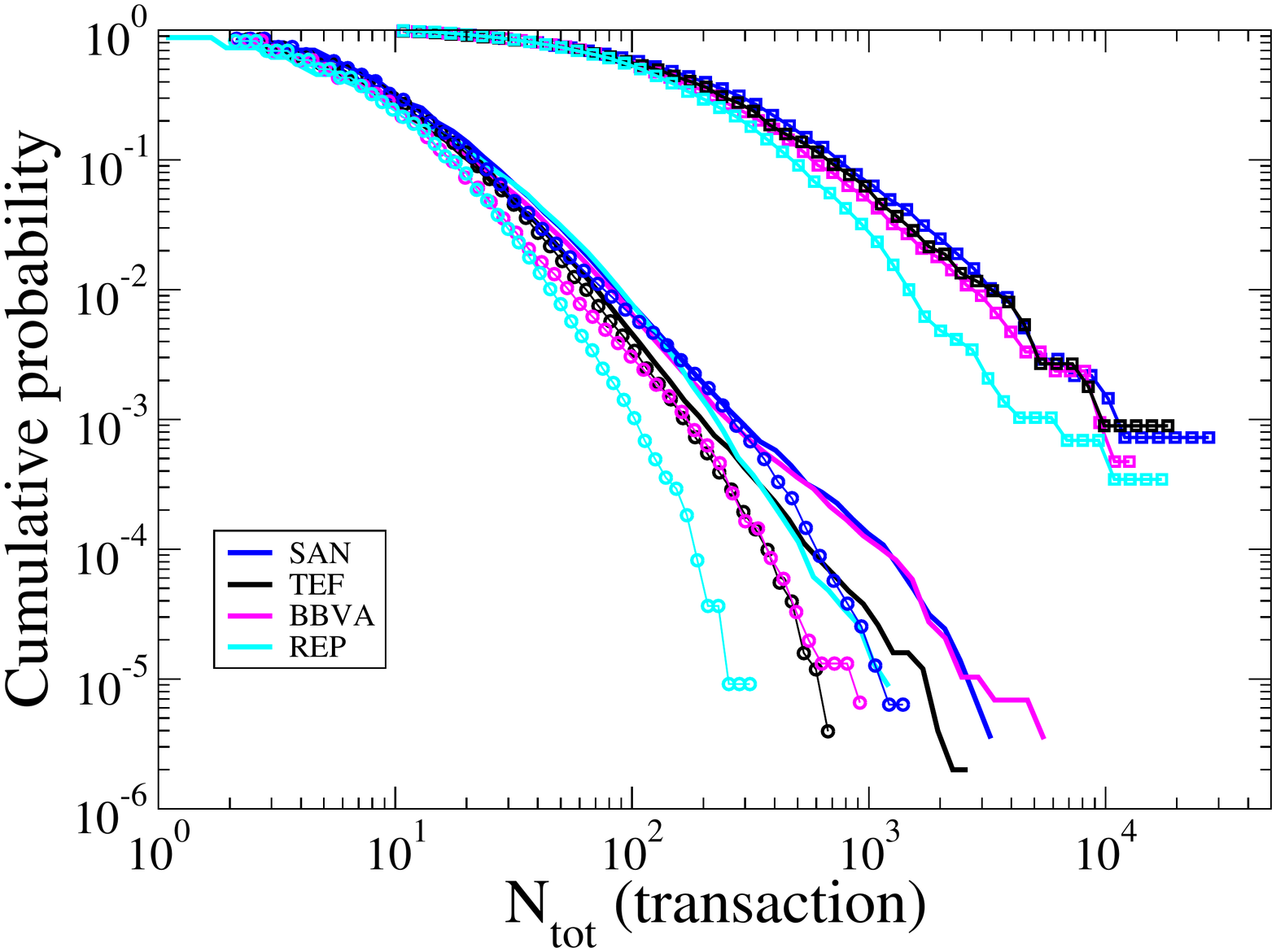}
  \includegraphics[width=8cm]{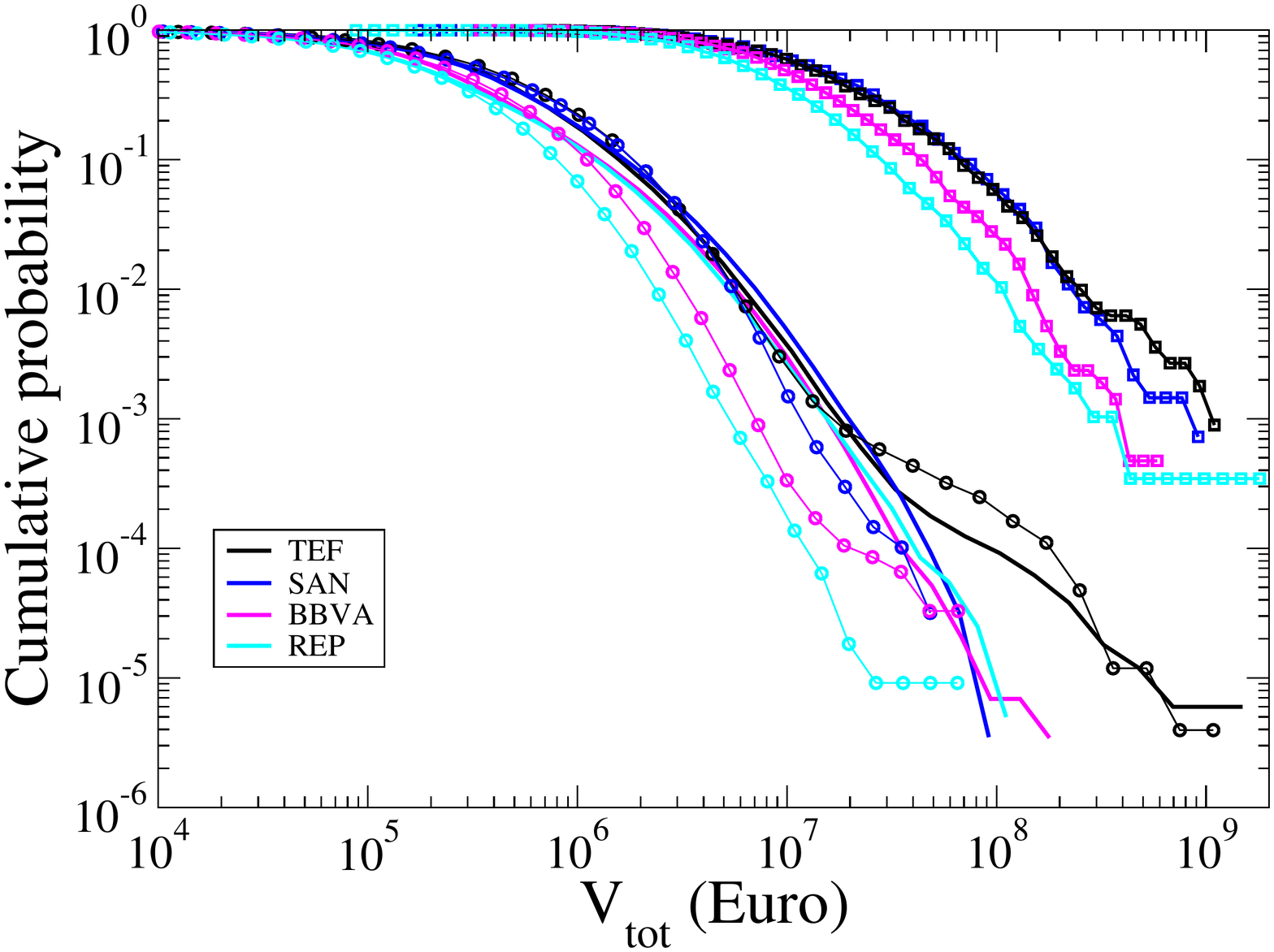}
 \caption{Top panel. Distribution of time interval $T$  elapsed between the first and the last transaction in an HMM patch.  Middle panel. Distribution of the total number of transactions $N_{tot}$ in an HMM patch. Bottom panel. Distribution of total total volume $V_{tot}$ exchanged in an HMM patch.  The solid lines refer to directional patches and the circles to the neutral patches. The squares show the distribution of the $T$, $N_{tot}$, and $V_{tot}$ variables for patches detected with the algorithm of Ref. \cite{Vaglica2008}. The investigated years are 2003-2006.}
 \label{distribHMM}
 \end{figure}

Three variables characterize the size of patches: $T$, $N_{tot}$, and $V_{tot}$. The time $T$ measured in seconds is the time elapsed from the first to the last transaction of the patch. We use trading time, i.e. we remove the overnight, the holidays, and in general any time when the market is closed. The total number of transactions composing the patch  is given by $N_{tot}=N_{buy}+N_{sell}$ and the total volume traded in the patch is $V_{tot}=V_{buy}+V_{sell}$. These variables are different from the one used in Ref. \cite{Vaglica2008} where only the most frequent type  of transactions (buy or sell) in a directional patch were considered in the definition of $N$ and $V$. However, as we have seen above, in HMM directional patches the most frequent type of transactions is responsible of almost all the volume exchanged in the patch and in fact we observe very little difference in the distributions obtained by considering all the transactions or only the ones of most frequent type. The motivation for the slight change in variables is that our present choice allows to investigate the distributional properties both of the directional and of the neutral patches with the same set of variables.

\begin{table}
\centering
\begin{tabular}{r||c|c|c|c|}
 &BBVA&REP&SAN&TEF\\
\hline
$\zeta_{T}$&$1.33 \pm 0.02$&$1.50 \pm 0.03$&$1.40 \pm 0.02$&$1.22 \pm 0.01$\\
$\zeta_{N_{tot}}$&$1.95 \pm 0.03$&$1.93 \pm 0.03$&$1.84 \pm 0.03$&$1.92 \pm 0.02$\\
$\zeta_{V_{tot}}$&$1.73 \pm 0.03$&$1.72 \pm 0.03$&$1.75 \pm 0.02$&$1.82 \pm 0.02$\\
\hline
$\zeta_{T}$&$1.31 \pm 0.03$&$1.34 \pm 0.03$&$1.36 \pm 0.03$&$1.26 \pm 0.02$\\
$\zeta_{N_{tot}}$&$2.07 \pm 0.04$&$2.48 \pm 0.06$&$1.87 \pm 0.03$&$2.19 \pm 0.03$\\
$\zeta_{V_{tot}}$&$2.36 \pm 0.04$&$2.31 \pm 0.05$&$2.38 \pm 0.05$&$2.15 \pm 0.03$\\
\end{tabular}
\caption{\footnotesize{Tail exponent of the cumulative distributions of $T$, $N_{tot}$, and $V_{tot}$ of HMM patches. The exponent is estimated with the Hill estimator applied on the top $5\%$ quantile. The top part of the table refers to directional patches (buy and sell state), and the bottom part refers to the neutral state. The errors identify a $95\%$ confidence interval.}}
\label{Hill-HMM5}
\end{table} 
\begin{table}
\centering
\begin{tabular}{r||c|c|c|c|}
&BBVA&REP&SAN&TEF\\
\hline
$\zeta_T$& $1.53 \pm0.25$& $1.73 \pm 0.24$&$1.54 \pm 0.32$&$   1.23 \pm0.28$\\
$\zeta_{N_{tot}}$&$1.64 \pm0.27$&$   2.10 \pm0.30$&$   1.57 \pm0.32$&$   1.58 \pm0.36$\\
$\zeta_{V_{tot}}$&$1.91 \pm0.31$&$   1.83 \pm0.26$&$   2.20 \pm0.45$&$   1.79 \pm0.41$\\     
\end{tabular}
\caption{\footnotesize{Tail exponent of the cumulative distributions of $T$, $N_{tot}$, and $V_{tot}$ of patches. The exponent is estimated by the Hill estimator applied on the top $5\%$ quantile of the directional patches detected in the signed volume time series by using the segmentation algorithm of Ref. \cite{Vaglica2008}. The errors identify a $95\%$ confidence interval.}}
\label{Hill-segment}
\end{table}

Figure \ref{distribHMM} shows the cumulative distribution for $T$, $N_{tot}$, and $V_{tot}$ for the four considered stocks pooling together the four different years. We pool together the directional patches because in this case we find a negligible difference of behavior between buy and sell patches. This is true also when we investigate the 2001-2002 period. We observe that for neutral patches, the distributions of $N_{tot}$ and $V_{tot}$  have a thinner tail than the corresponding directional patch distributions. For $T$ the distributions for the three states are very similar. The cumulative distribution shows that approximately $90\%$ of the detected patches are characterized by a time duration of less than an hour. Therefore the largest fraction of detected patches seems to be proxies of intraday hidden orders. This might indicate a preference of the traders to close daily the position taken or might be related to a specific aspect of our approach which is monitoring market members and not single investors. In fact it is realistic to hypothesize that market members are most of the time performing multiple, although heterogeneous, strategies and the interference of multiple strategies might end up into a local fragmentation of trading strategies planned over longer time horizons by a limited number of large investors. Fragmentation which is revealed by the HMM approach.    

However in spite of this limitation, from the figure \ref{distribHMM} we can see that also in the case of HMM, patch size has a fat tail distribution. We performed a Jarque-Bera test of the hypothesis that the above distributions are lognormal and in all the cases we reject the null hypothesis $0.01$ confidence level. It is difficult to assess if the tail of the distributions is power law given the small extension of the tail. In order to compare the tail of the distributions with those obtained in Ref. \cite{Vaglica2008} we estimate the tail exponent by using the Hill estimator.  Table \ref{Hill-HMM5}  gives the Hill estimation for the tail exponent for all the analyzed sets and variables when the estimation is applied to the top $5\%$ percentile.

We also compare the size distribution of the HMM patches with those found by applying the segmentation method used in Ref. \cite{Vaglica2008} to the signed volume time series. We therefore run the segmentation algorithm to the inventory time series and we consider as size variables $T$, $N_{tot}$, and $V_{tot}$.
The figure shows that these patches are much longer than the one identified by the HMM.  Table \ref{Hill-segment} shows the Hill estimator of the tail exponent of the directional patches identified by the segmentation algorithm on the top $5\%$ quantile. By comparing the exponents in this table and in Table \ref{Hill-HMM5} we observe that the values are not very different suggesting a relation between the small HMM patches and the larger patches detected by the segmentation algorithm. In Section 7 we will make this comparison more quantitative.

\section{The buy-sell asymmetry}\label{20012002}

\begin{figure}[ptb]
\begin{center}
\includegraphics[scale=0.35]{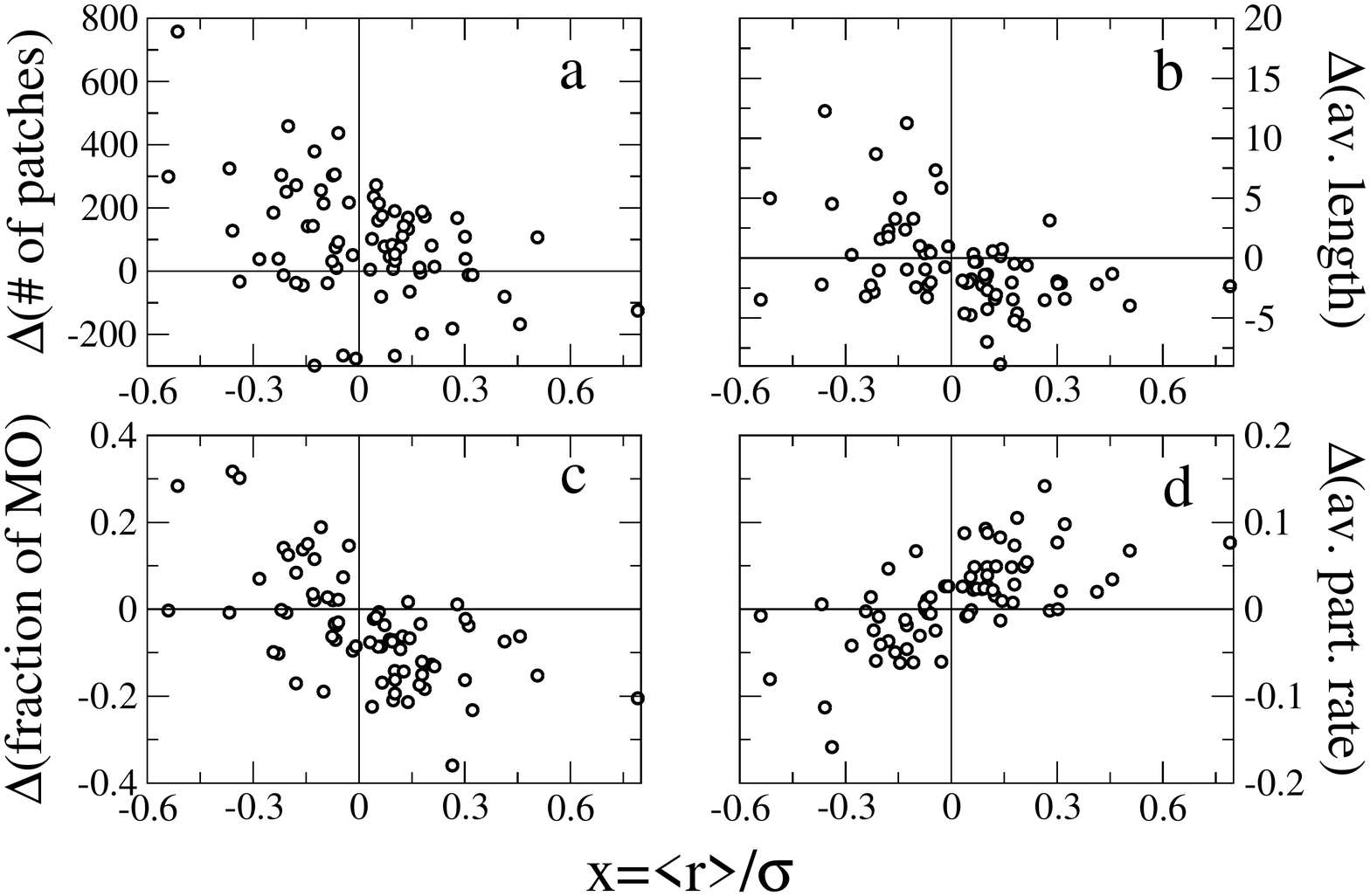}
\includegraphics[scale=0.35]{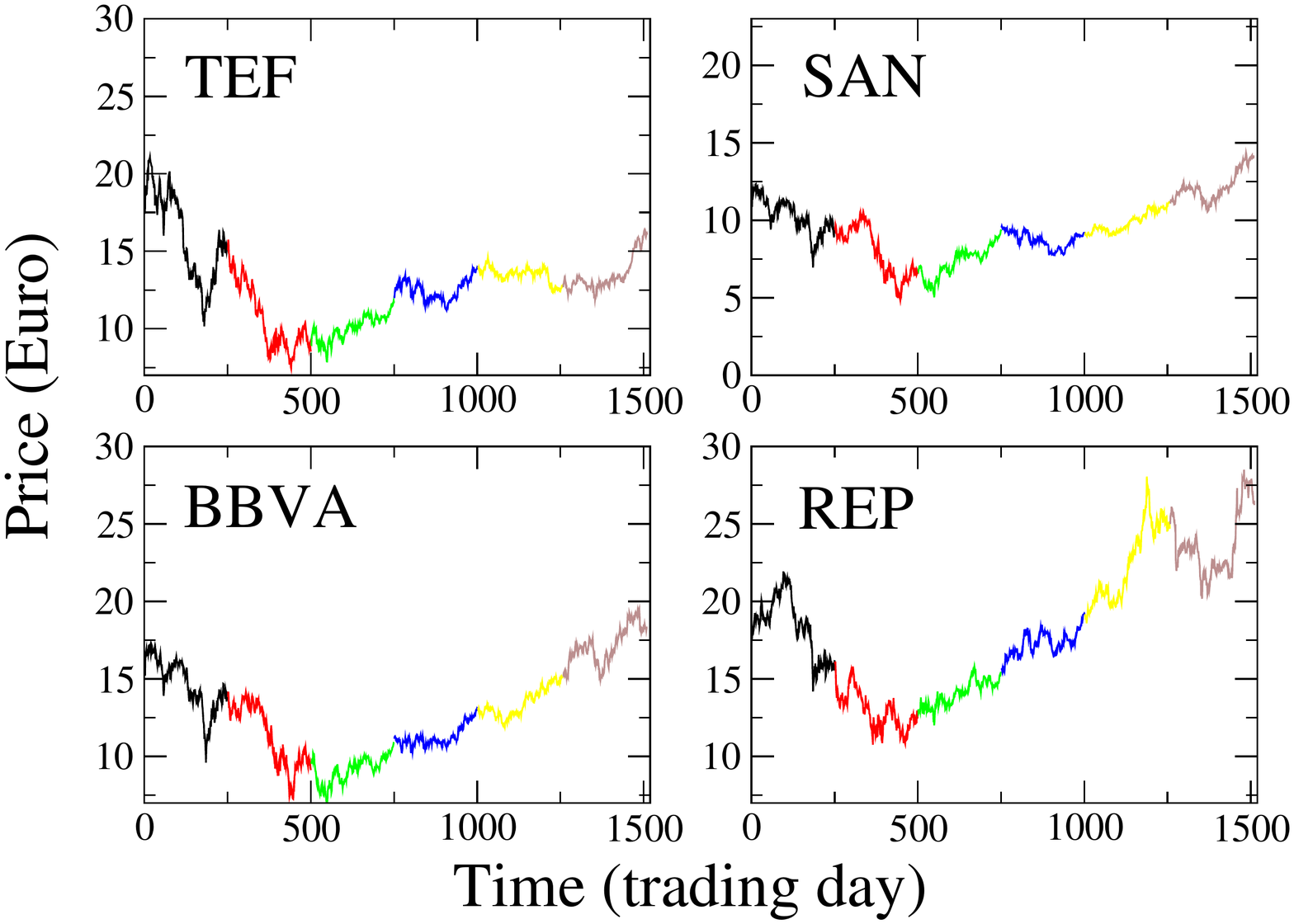}
\end{center}
\caption{Top 4 panels. Buy-sell asymmetry measures as a function of the ratio $x=<r>/\sigma$ for directional patches with at least 10 transactions. The symbol $\Delta(.)$ indicates the difference of the quantity in parenthesis between buy and sell patches. Specifically, panel (a) shows the difference of the number of buy and sell patches, panel (b) shows the difference of the average length (in transactions) between buy and sell patches, panel (c)  shows the difference of the average fraction of market orders between buy and sell patches, and finally panel (d) shows the difference of the average participation rate between buy and sell patches. Bottom 4 panels. Closing price time series for the four investigated stocks in the period 2001-2006. Each color refers to a different calendar year.}
\label{price}
\end{figure}

We have seen in the previous sections that several properties of HMM patches show a significant behavior asymmetry between buy and sell patches. This may be due to two possible reasons. The first possible explanation is that there is an intrinsic asymmetry between the buy and sell strategy of hidden order splitting. This may be due to behavioral biases, margin calls, short selling restriction, etc. The second possible explanation is related to the presence of a different behavior in the buy-sell order splitting strategy which is {\it conditional} to the market state, for example the fact that the price is trending up or down. In this second case there is a buy-sell asymmetry when the market is, for example, trending up, but the asymmetry is reverted when the market trends down.

In order to investigate which hypothesis explains better the data, we consider Telefonica stock and we divide our six year period (2001-2006) in 72 one month sub-periods. For each of them we compute the ratio $x=<r>/\sigma$ between the average daily price return in the month and the volatility $\sigma$, here computed as the standard deviation of daily returns in the month. This ratio gives a measure of the price trend in the month. Then we consider the HMM directional patches detected in the considered month with at least 10 transactions and we compare the statistical properties of the buy and sell patches as a function of  $x$. The top part of Figure \ref{price} shows the results. Specifically, panel (a) shows the difference of the number of buy and sell patches as a function of $x$. Panel (b) shows the difference of the average length measured in transactions between buy and sell patches  as a function of $x$. Panel (c) shows the difference of the average fraction of market orders between buy and sell patches as a function of $x$, and finally panel (d) shows the difference of the average participation rate between buy and sell patches as a function of $x$. In each panel we observe a highly significant linear relation between the plotted variables. The correlation coefficient is -0.42, -0.42, -0.61, and 0.64, respectively. This result indicates that when the price increases there are more sell patches, which are typically longer, richer of market order, and with a lower participation rate when compared to buy patches. The opposite happens when the price decreases. The intercept of the regressions is significantly different from zero for the number and the fraction of market orders suggesting also an overall buy-sell asymmetry even in the absence of a price trend ($x=0$). However the dependence of the four quantities from $x$ is much stronger, indicating that the second hypothesis stated above is probably the most relevant one in explaining the buy-sell asymmetry. Finally, for the neutral patches we do not observe any significant dependence of the statistical properties as a function of $x$. A full explanation of these asymmetries is beyond the scope of this paper and will be considered in a future paper.

These asymmetries help to explain why we investigate the market in 2003-2006 and leave out the period 2001-2002.  The bottom part of Figure \ref{price} shows the closing price of the four stocks in the period 2001-2006. We see that the market was in a bear phase (i.e. decreasing price) in 2001-2002, while it was in a bullish phase (i.e. increasing price) in 2003-2006. By investigating each year separately we find that some properties of the HMM patches are similar in all years, while others have a different behavior in 2001-2002 and in 2003-2006. The main difference we find in 2001-2002 are:
\begin{itemize}
\item{Sell patches have typically a smaller fraction of market orders than buy patches, i.e. the contrary of what observed in 2003-2006 where sell patches have a higher fraction of market orders than sell patches. (figures \ref{mopdf} and \ref{length-mo})}
\item{Sell patches have a higher participation rate than buy patches, i.e. the contrary of what observed in 2003-2006 where sell patches have an smaller participation rate than buy patches (figures \ref{aggressiveness-pdf} and \ref{aggressiveness}).}
\end{itemize}

These differences are consistent with the dependence of the buy-sell asymmetry from the price trend investigated above on the monthly scale. We observe that this explanation seems to work even when one considers longer time scales such as the yearly time scale. While this explains our choice of investigating only the four year period 2003-2006, a full understanding of the dependence between long scale price trends and short scale HMM buy-sell asymmetries requires  further investigations.

\section{Comparison of segmentation algorithms}\label{Comparison}

\begin{figure}[ptb]
\begin{center}
\includegraphics[scale=0.27]{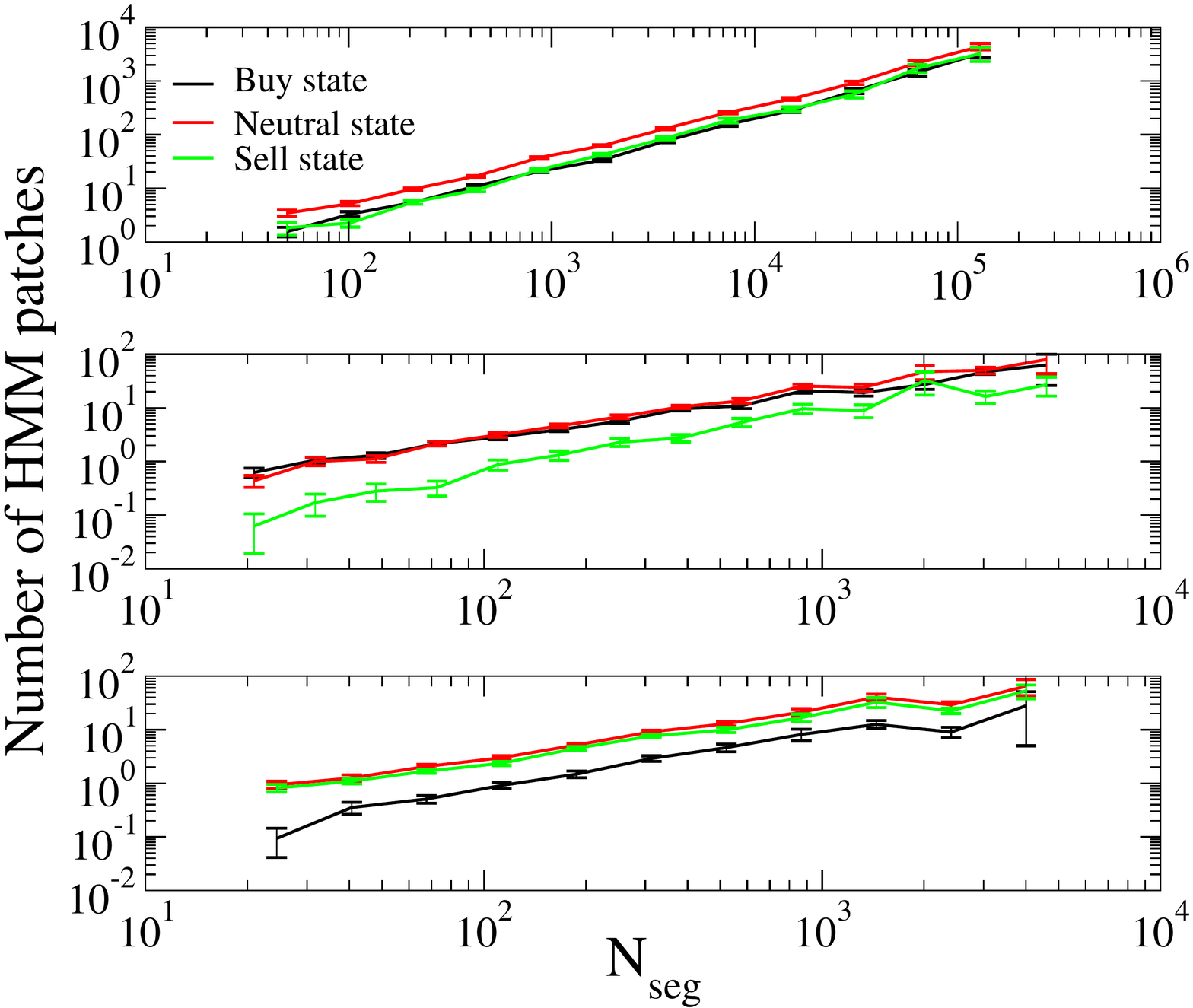}
\includegraphics[scale=0.27]{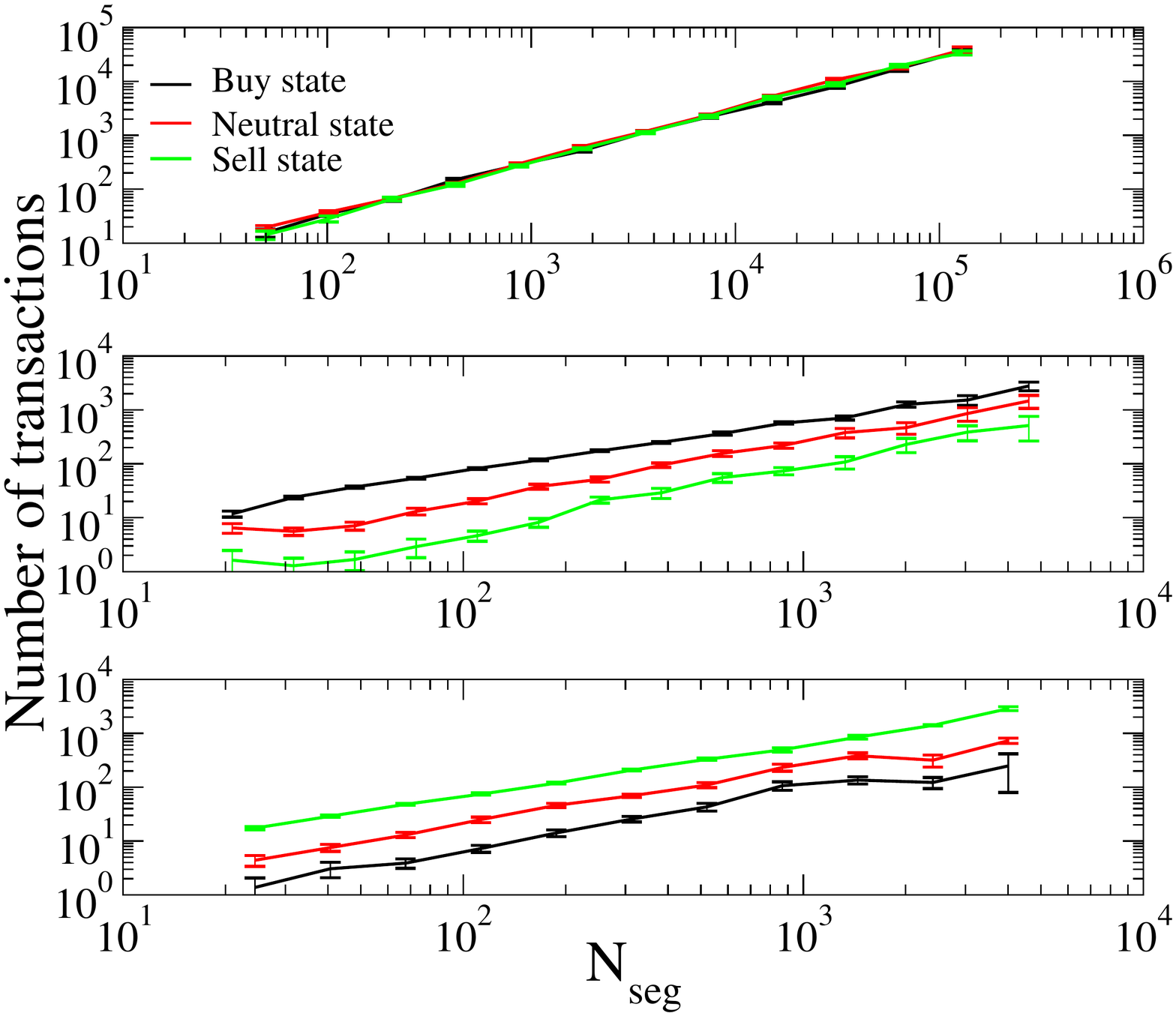}
\end{center}
\caption{Left panel. Average number of HMM patches in the three HMM states found inside a segment patch of type neutral (top), buy (middle) and sell  (bottom) as a function of $N_{seg}$, which is the length of the segment patch. Right panel. Average number of transactions in HMM patches in the three HMM states found inside a segment patch of type neutral (top), buy (middle) and sell  (bottom)  as a function of $N_{seg}$.}
\label{comparison}
\end{figure}

In this section we compare the patches detected with the HMMs and the ones detected with the algorithm of Ref. \cite{Vaglica2008}  in more detail. As we have described in Section 6, the two types of patches describe regimes at very different scales. HMM patches are typically short, while the the patches of  Ref. \cite{Vaglica2008} (hereafter addressed as {\it segment patches}) are very long. In order to compare the two segmentations, we divide the segment patches in three groups of type buy, sell, and neutral according to the criterion used in Ref. \cite{Vaglica2008}. We therefore compute the number of HMM patches in the three HMM states contained in a segment patch of a given type. The left panel of Figure \ref{comparison} shows the average number of HMM patches in the three HMM states for segment patches of the three types 
as a function of $N_{seg}$, which is the number of transactions present in the segment patch. Neutral segment patches contain roughly the same number of neutral HMM patches as the sum of the number of buy and sell HMM patches. The middle and bottom panels of the left part of figure  \ref{comparison} show that a buy (sell) segment patch contains typically an equal number of buy (sell) and neutral HMM patches and a much smaller number of sell (buy) HMM patches. Therefore directional segment patches are typically a mixture of neutral and directional (with the same sign) HMM patches. 

The comparison is even clearer if one considers the number of transactions in a segment patch, which are assigned by the HMM to a specific type of state (see the right panel of Fig. \ref{comparison}). Neutral segment patches have on average an equal number of transactions in the three HMM states (see the top right panel). Taken together with the top left panel, this result indicates that neutral segment patches are composed by relatively short neutral HMM patches and an equal mixture of longer buy and sell HMM patches. By contrast buy (sell) segment patches are composed by a large fraction of transactions in the buy (sell) HMM state, a smaller fraction of transactions in the HMM neutral state, and a much smaller fraction of transaction in the opposite directional HMM state. We therefore conclude that directional segment patches are mainly composed by an equal mixture of short neutral HMM patches and long HMM directional patches with the same direction as the segment patches.   

\section{Conclusions}

In this paper we have shown that HMMs can be used to detect patches in the series of transaction signs performed by market members. We interpret the directional patches as possible hidden orders submitted by the market members. By analyzing the statistical properties of these patches we observe that the time distribution of these patches is fat tailed.  Unconditionally, long patches are characterized by an high fraction of market orders and a low participation rate, while short patches have a large fraction of limit orders and an high participation rate. We also observe an interesting buy-sell asymmetry which is depending on the trend of the stock price. When the price goes up, we detect more sell patches, which are longer, with more market orders, and less aggressive compared to the buy patches detected in the same period. When the price decreases the opposite behavior is observed.  These results may be considered a contribution in the identification of stylized facts in the strategic behavior of the investors.

The comparison of the HMM patches and those obtained with the segmentation algorithm used in  \cite{Vaglica2008} can shed light on the their interpretation.
The two segmentation methods perform a coarse graining of the inventory time series on two different time scales. The HMM is able to detect short patches with a well defined drift, while the segmentation algorithm  detects large scale patches composed by many smaller HMM directional patches with the same sign interspersed by even smaller HMM neutral patches. A possible interpretation is that the segment patches represent trading decision of the portfolio managers that are executed on  time scales that can go up to several days, while HMM patches represent trader's executions of usually intraday smaller packages, which might be part of the larger portfolio manager orders.

There are many interesting points raised by the present study that might deserve future investigations. First, it is important to build a stronger relation between the stylized facts observed in HMM patches and the possible strategies adopted by the investors. Second, the buy-sell asymmetry deserves a more in depth analysis to understand (i) how it depends on the considered time scale, and (ii) a possible causal relation between the asymmetry and the price trend. Specifically, is the price going up or down because of the asymmetry or is the order flow adapting to the price drift? Third, it is interesting, both from a theoretical and from an applied point of view, to compute the market impact of the HMM patches, in a way similar to what has been done in Ref. \cite{Moro2009} for patches detected with the segmentation algorithm. Finally, in our study we have pooled together all the market members without considering possible differences in the order splitting strategies across market participants. The understanding of the heterogeneity and mutual relationship in the order splitting strategies of market participants might help in understanding the fascinating process of price formation.

\section*{Acknowledgments} Authors acknowledge financial support from the PRIN project 2007TKLTSR  ``Indagine di fatti stilizzati e delle strategie risultanti di agenti e istituzioni osservate in mercati finanziari reali ed artificiali".

\appendix
\section{~~~~~~~~~~~~ Hidden Markov models, Hidden Semi Markov models and power law distributions}\label{appendix}

In this paper we have used HMMs to model hidden orders. However HMMs are parametric models and therefore assume a given structure of the data. A known property of HMMs is that patch length are exponentially distributed. In fact, if a system described by an HMM , which is initially in a given state $S_i$, the probability $p_i(\ell)$ that the system stays in that state for exactly $\ell$ steps, i.e. that
\begin{eqnarray}
O=\{q_1=S_i,q_2=S_i,q_3=S_i,q_4=S_i,...,q_\ell=S_i,q_{\ell+1}\neq S_i\}
\end{eqnarray}
is
\begin{eqnarray}
p_i(\ell)=(a_{ii})^{\ell-1}(1-a_{ii}) \label{t_steps}
\end{eqnarray}

Since we do not have any prior on the length distribution, we test numerically whether HMMs are able to detect non-exponential distributed patch lengths. More specifically, we perform two in-depth simulation studies. In the first we simulate an artificial time series with patch length taken from a Pareto (i.e. power law) distribution. Then we fit an HMM and we reconstruct the hidden states of the process by looking for the best sequence of states, which are individually most likely.
The second simulation study consists in considering a generalization of HMMs, namely the hidden semi Markov model (HSMM) in which the patch length distribution is not fixed by the model but it is fitted from the data. Both studies show that our procedure is able to detect non exponential distributed patch lengths.

\subsection{Reconstruction of power law distributed patches}\label{patches}
\begin{figure}[ptb]
 \begin{center}
  \includegraphics[width=8cm]{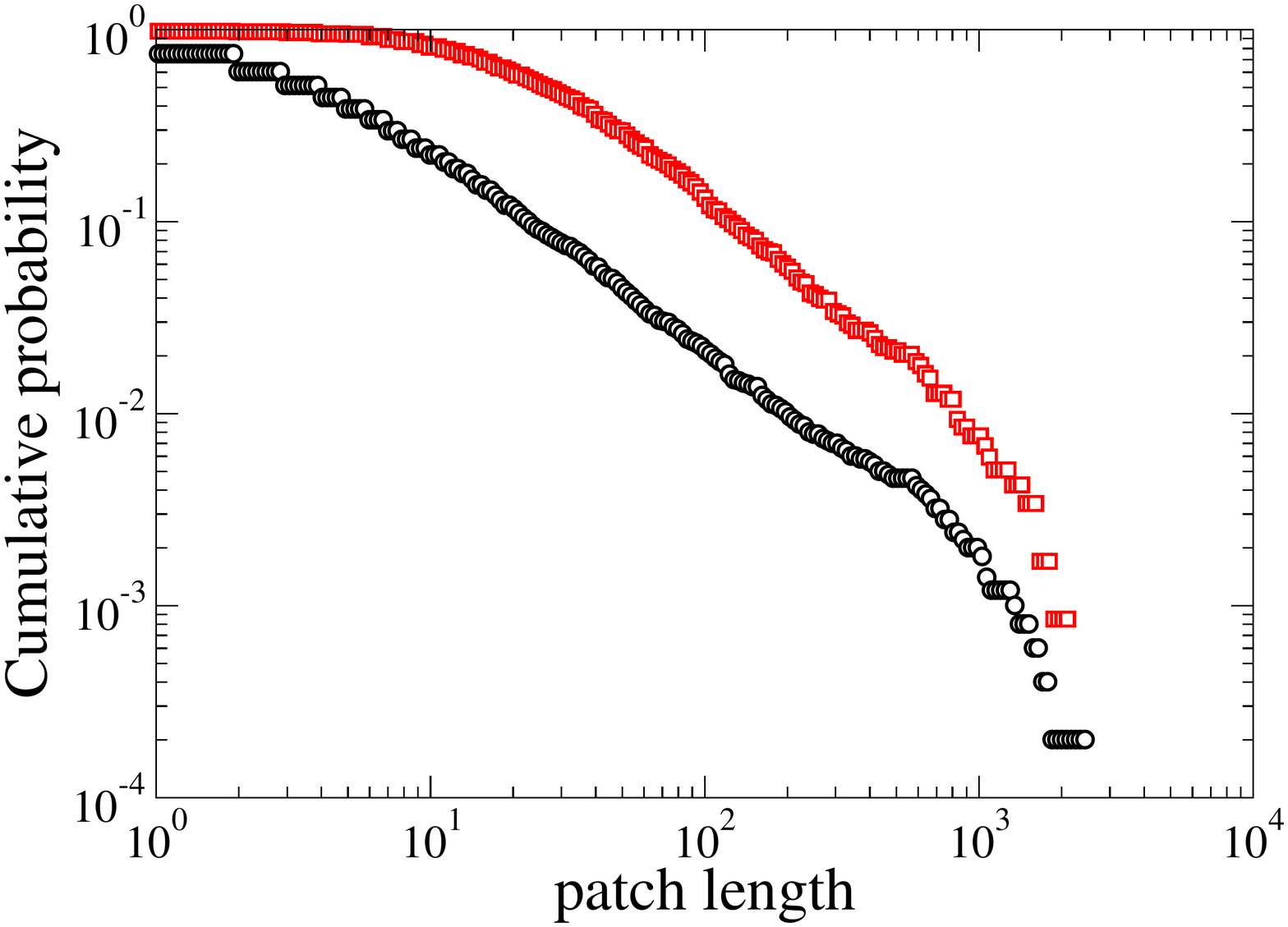}
 \end{center} 
  \caption{\footnotesize{Black circles. Cumulative probability of the patch length simulated from a Pareto distribution of tail exponent $\mu=2$.
                         Red squares. Cumulative probability of the patch length detected by the HMM}}
  \label{ParetoHmm}
\end{figure}

Here we show the results of a simulation study of the reconstruction with HMM of the hidden states of an artificial time series generated with power law distributed patch length. 
Specifically, we simulate time series composed by patches of variable length and characterized by different composition of the symbols $+1$ and $-1$. The length of each patch is
extracted from a Pareto distribution $P(\ell)\propto \ell^{-\mu}$. We assign to each patch a dominant sign ($+1$ or $-1$), and then for each patch we simulate a time series of length $\ell$ as a binary time series with a bias in favor of the dominant sign. Consecutive patches have alternate dominant signs. In the following we show results obtained for a surrogate time series composed by 5,000 patches with length  extracted from a Pareto distribution with exponent $\mu=2$. In figure \ref{ParetoHmm} we compare the distribution of the simulated patch length  with the distribution of the HMM reconstructed patch length. Figure \ref{ParetoHmm} shows that the tail behavior of the two distributions is quite similar, a part from a global factor which is due to the fact that the HMM is unable to detect very short patches due to statistical fluctuations. This result suggests that HMMs are able to reproduce the power law behavior of the distributional properties of the patches in a time series, despite the exponential structure of the patch length expected in an HMM.

\subsection{~~~~~~~~~~ Comparison of hidden Markov models and hidden semi Markov models}\label{hmms}

\begin{figure}[ptb]
\begin{center}
\includegraphics[scale=0.55]{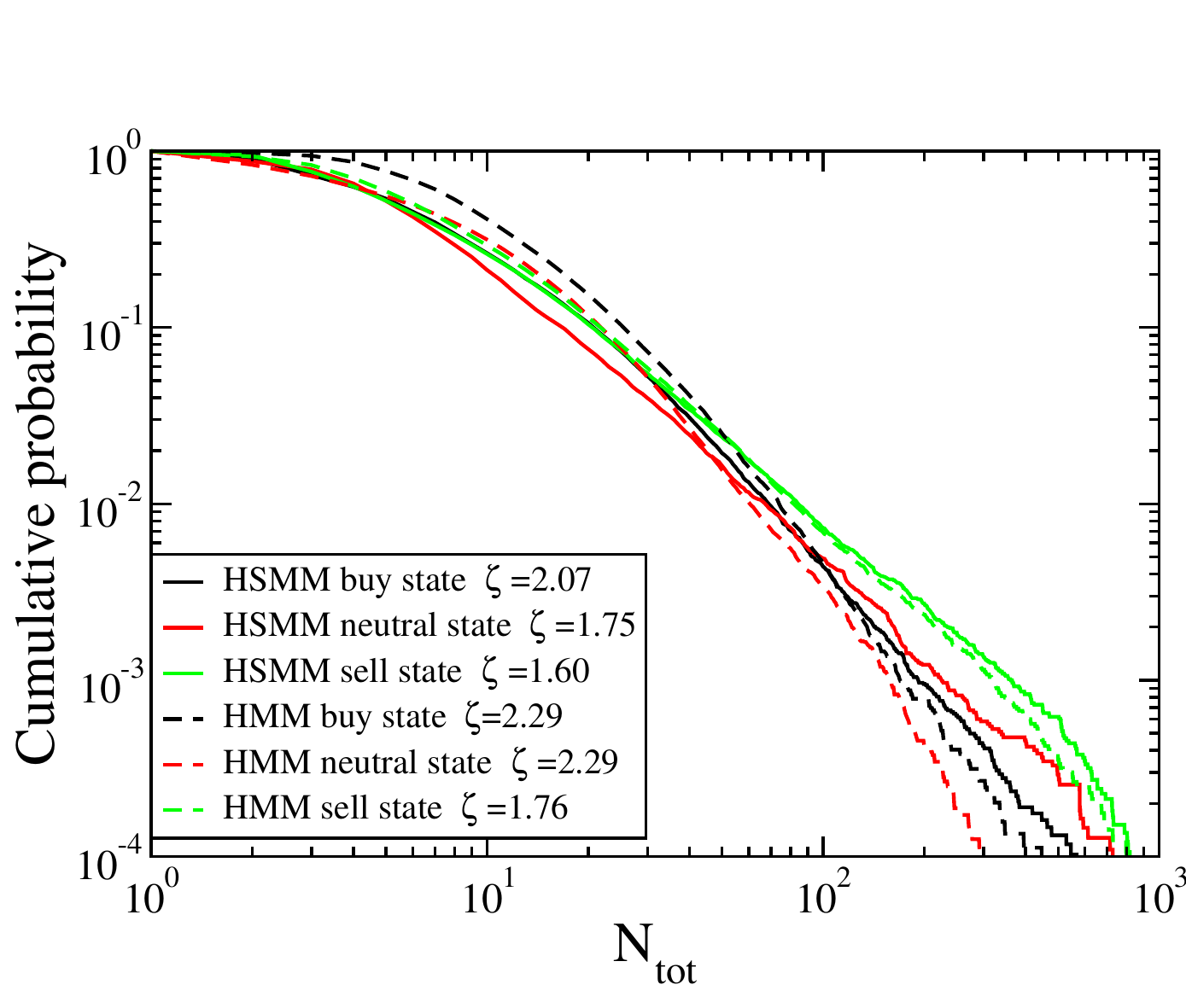}
\end{center}
\caption{Cumulative probability of the patch length for the three states obtained with the HSMM (solid line) and with the HMM (dashed lines). In the legend there are also shown the respective value of tail exponent $\zeta$ performed by Hill estimation on the top  5\% quantile.}
\label{distribHSMM}
\end{figure}

The problem of the exponential distribution of the patch length is well known in the HMM literature. The most used extension of HMMs to cope with the problem of exponential distribution of patch length is the HSMM  \cite{Ferguson1980}. In an HSMM the diagonal elements of the transition probability are set to zero and the model is specified by assigning also the patch length distribution (sometimes called sojourn time distribution, or runlength distribution) associated with each state and defined as
\begin{equation}
d_j(\ell)=P(S_{t+\ell+1}\ne j, S_{t+\ell}=j,...,S_{t+2}=j|S_{t+1}=j, S_t\ne j)  \hskip0.2cm 1\leq j \leq N .
\end{equation}
In other words in an HSMM the patch length distribution of each hidden state is assigned explicitly in the specification of the model and is not therefore necessarily exponential. There exist algorithms that allows to fit an HSMM with a given sojourn time distribution to a time series. However if one has no prior about the functional form of the sojourn time distribution, it is possible to fit a non parametric HSMM to a time series. In this case the distributions of sojourn time are fitted from the data.  HSMMs with nonparametric state sojourn time distributions were first proposed by Ferguson in 1980 \cite{Ferguson1980} in the field of speech recognition and an efficient algorithm for the model fit has been recently proposed \cite{Guedon2003}.  In the HSMM investigation we use a non parametric fitting, i.e. we fit also the distribution of sojourn time. This estimation requires long  computational time and can be only applied to long time series. 

For the estimation of the model parameters of the HSMM and the subsequent reconstruction of the underlying state sequence we use of the  $R$ library \emph{hsmm} \cite{Bulla2008}. The algorithms for maximum likelihood parameter estimation in this library are based on a method introduced in \cite{Guedon2003}. 
The non parametric estimation of the sojourn time distribution requires a high number of observations. For this reason we choose to analyze only time series with at least 20,000 records. Accordingly we select 28 market members trading TEF in the 2006. 
The mean  buy probability computed across market members in the three hidden states of the fitted HSMM is $0.05 \pm 0.05$, $0.53 \pm 0.16$, and $ 0.97 \pm 0.05$, which supports the interpretation of the three states as sell state , neutral state , and buy state, respectively. The mean transition probability matrix is
\begin{eqnarray}
\left(\begin{array}{ccc}
0&0.3\pm0.2&0.7\pm0.2\\
0.4\pm0.2&0&0.6\pm0.2\\
0.6\pm0.1&0.4\pm0.1&0\\
 \end{array}\right)
\end{eqnarray}

After the model parameter estimation, we compute the hidden state sequence for each of the 28 inventory variation sign time series. 
In figure \ref{distribHSMM} we compare the cumulative probability of patch length obtained with the HMM and the HSMM. For the directional states the distribution of patch length obtained with the two methods are quite similar. The most pronounce difference is observed in the tail of the distribution of length of neutral patches. We also estimate the tail exponents with the Hill estimator on the 5\% quantile and we observe similar tail exponent values for HMM and HSMM patches.

The parameter estimations for the HSMM requires much more computational resources than the HMM. The fitting and reconstruction the time series for one year of transactions of 28 participants with HSMM required about 6 days of computing time against a few minutes required for the HMM modeling. The complexity of the algorithm is linear in the sequence length, and we estimated that the modeling of 6 years of transactions for the participants trading the stock Telefonica would have required roughly one month of computations. Even if  HSMM is a more realistic choice than HMM in modeling inventory time series, the small differences in the distributions observed in Fig. \ref{distribHSMM} convinced us to use the HMM  to model our data.

\end{document}